\documentclass[12pt,preprint]{aastex}

\newcommand{\hi}{{H~{$\scriptstyle {\rm I}$}}}

\begin{document}
\title{A Very Sensitive 21cm Survey for Galactic High-Velocity \hi}
\shorttitle{High-Velocity \hi}
 
\author{Felix J. Lockman}
\affil{National Radio Astronomy Observatory
\footnote{The National Radio Astronomy Observatory is a facility of the
National Science Foundation operated under cooperative agreement with
Associated Universities, Inc.}, P.O. Box 2, Green Bank, WV, 24944; 
 jlockman@nrao.edu}
\authoraddr{P.O. Box 2, Green Bank, WV, 24944}
\author{Edward M. Murphy}
\affil{University of Virginia, Department of Astronomy,
P.O. Box 3818 University Station, Charlottesville, VA 22903-0818;
emm8x@virginia.edu}
\authoraddr{Department of Astronomy, University of Virginia
P.O. Box 3818 University Station, Charlottesville, VA 22903-0818
emm8x@virginia.edu}
\author{Sara Petty-Powell}
\affil{ The Evergreen State College, Physics and Astronomy,
2700 Evergreen Parkway NW, Olympia, WA 98502}
\authoraddr{Physics and Astronomy, The Evergreen State College, 
2700 Evergreen Parkway NW, Olympia, WA 98502}
\and
\author{Vincent J. Urick \altaffilmark{2} 
\altaffiltext{2}{Current address:  
Naval Research Laboratory, Optical Sciences, Code 5652, 
Building 215, Room 120, 4555 Overlook Ave. SW, Washington, D.C. 20375}}
\affil{Bloomsburg University, Physics Department, 400 East 2nd St.,
Bloomsburg, PA 17815}
\authoraddr{Naval Research Laboratory, Optical Sciences, Code 5652, 
Building 215, Room 120, 4555 Overlook Ave. SW, Washington, D.C. 20375}

\begin{abstract}

Very sensitive \hi\ 21cm observations have been made in 860 directions
 at  $\delta \geq -43\arcdeg$  
in search of weak, Galactic,  high-velocity \hi\ emission lines
at moderate and high Galactic latitudes.
One-third of the observations were made toward extragalactic 
objects that are visible at optical and UV wavelengths. 
The median rms noise in the survey spectra is 3.4 mK, 
resulting in a median $4\sigma$ detection level of 
$N_{HI} = 8 \times 10^{17}$ cm$^{-2}$ averaged over the $21 \arcmin$ 
beam of the telescope.  
High-velocity \hi\ emission is detected in 37\% of the 
directions; about half of the lines could not have been detected in 
previous surveys.  The median FWHM of detected lines is 30.3 km~s$^{-1}$. 
High-velocity \hi\ lines are seen down to the sensitivity limit of the 
survey implying that there are likely lines at 
 still lower values of $N_{HI}$.  The weakest lines have a 
kinematics and distribution on the sky similar to that of the 
strong lines, and thus do not appear to be a new population.   
Most of the emission originates from objects which are extended 
over several degrees; only a few  appear to be 
compact sources.  At least 75\%, and possibly as many as 90\%, 
 of the lines are associated with one of the major high-velocity
complexes.  With the increased sensitivity of this survey, 
 the Magellanic Stream is seen to extend 
at least  $10\arcdeg$ to  higher Galactic latitude than previously
thought and to be more extended in longitude as well. 
Wright's Cloud near M33 has an extended low-$N_{HI}$ component in the 
direction of the Magellanic Stream. 
The bright \hi\ features which have dominated most surveys may be 
mere clumps within larger structures, and not independent objects. 
Although there are many lines with low column density, 
their numbers do not increase as rapidly as $N_{HI}^{-1}$, so most of the \hi\ 
mass in the high-velocity cloud phenomenon likely resides 
in the more prominent clouds.

\end{abstract}
 

\section{Introduction}

Emission in the 21 cm \hi\ line is seen over a large fraction of the sky
at velocities $-500 \lesssim V_{LSR} \leq -100 $ km~s$^{-1}$ and 
$+100 \leq V_{LSR} \lesssim +400 $ km~s$^{-1}$: too large to 
 arise solely from Galactic rotation, yet seemingly not part of the 
Hubble flow  (see 
\citet{wvw97} for a recent review; throughout this work we use the
 definition that high-velocity \hi\ has $|V_{LSR}|\geq100$ 
km~s$^{-1}$ in directions
where such velocities are not expected from Galactic rotation). 
These  high-velocity clouds were
discovered in the 21cm line of \hi\  \citep{muller}, and for many decades 
the 21cm line was their primary spectral signature.
Surveys of the sky in search of high-velocity \hi\ are 
thought to be complete at the level of  $N_{HI} \geq 2 \times 10^{18}$ 
cm$^{-2}$  for objects with an angular size of a degree or 
greater \citep{wakk91}.  The most 
prominent southern high-velocity clouds can be 
associated with the Magellanic Clouds \citep{mathewson}, 
  and some high-velocity clouds lie in the Galactic halo 
\citep{vanwoerden99}, 
but the origin of most of the  emission is 
unknown. It has been suggested that  high-velocity clouds 
result from phenomena as diverse as  extensions of the 
Galactic disk, material accelerated by supernovae or other energetic events, 
the remnants of a  Galactic fountain,  condensations in a Galactic 
halo, infall of debris from satellite galaxies, and  
objects in the Local Group of galaxies \citep{oort66, 
habing, davies, shapiro, bregman80, norikeu, blitz, kalkerp99, mallouris}.  
It is probable that high-velocity 
clouds are a heterogeneous population 
and  all of these possibilities are correct for 
different subsets of the data.  

A number of high-velocity absorption lines have been detected in the 
ultraviolet  toward extragalactic 
sources (e.g., \citet{savage93, bowen93, bowen95, Lu94, murphy00, gibson2000}).
 The UV transitions  are much more sensitive to small amounts of
neutral gas than the 21cm line: a cloud with $N_{HI} = 10^{18}$ cm$^{-2}$ 
will produce  strong Mg II absorption  even if the 
metallicity is only 0.1 solar \citep{ls95,charleton00}.  
The statistics of the UV absorption lines 
 suggest that high-velocity clouds are more abundant than 
would be expected from  21cm \hi\ surveys \citep{savsem96}. 
The extrapolation from the 21cm measurements to lower $N_{HI}$ levels 
is always uncertain  \citep{wakk91}, and the UV results are based
on relatively few sight lines, but these observations 
 raise the interesting 
 possibility that the Milky Way is surrounded by a mist of 
high-velocity gas with $N_{HI} \lesssim 10^{18}$ cm$^{-2}$ 
which has escaped detection in previous surveys.  This is
doubly interesting because the  HI disks of other spiral galaxies are 
truncated at  the level of a few $10^{19}$ cm$^{-2}$, presumably because of 
ionization  by the extragalactic radiation field 
\citep{vangorkom91, vangorkom93, bochk, maloney, corbelli}.  This 
possible population of clouds with low $N_{HI}$  might have a 
different distribution and different properties than the denser 
high-velocity clouds.  It might also contain significant mass.

To try to bridge the gap between the sensitivities of the 
21cm and the UV measurements, we began in 1992 
a very sensitive 21cm \hi\  search for high-velocity 
 lines too faint to have been
detected in previous 21cm surveys.  Some results from the
first third of the data have already been published \citep{mls95}.
Since  the beginning of this project there has been renewed interest 
in the high-velocity cloud phenomenon. 
 There has been an extensive 21cm survey of the southern 
high-velocity cloud population \citep{putmanetal2002}. 
There has been a proposal that 
most of the clouds are members of the Local Group at a distance $\sim 1$ Mpc
 \citep{blitz}, and another that  some fraction of them 
are dwarf irregular galaxies which lack bright stars \citep{bb99,bb00}. 
There have been 
investigations into the connection between 
high-velocity clouds and the neutral gas
in the local universe \citep{charleton00, zwaan00, zwann01}.
While the data presented here bear somewhat on these issues, 
the main goal of this work has always been
 to investigate the high-velocity cloud 
phenomenon  at low values of $N_{HI}$, and that aspect of the data
 is the focus of this paper.

\section{ Observations}

\subsection{Equipment}

     The observations were made with the 140 Foot (43m) Telescope of 
the NRAO in Green Bank, WV,  during nineteen observing sessions between  
March 1993 and September 1996.  The telescope
has an angular resolution of $21\arcmin$ at the wavelength 
of the 21cm \hi\ line. An HFET
receiver was operated in dual circular polarization mode and 
had a total system temperature at
zenith of about 20 K in each channel.  The spectrometer was the 
NRAO Model IV 3-level autocorrelator, which provided a 512 channel 
spectrum in each polarization.  Spectra were taken by 
position-switching using a 
 10 MHz bandwidth for a total velocity coverage $-1000 \leq V_{LSR} 
\leq +800 $ km~s$^{-1}$ and an effective velocity resolution of 
5.0 km~s$^{-1}$.  About 100 of the directions were reobserved   
using frequency-switching 
 at 2.5 MHz total bandwidth with an effective velocity resolution of 
1.25  km~s$^{-1}$ over $450$ km~s$^{-1}$.  
The  good velocity resolution  of the  spectra insures that
lines will always be present in several channels, 
minimizing the possibility that narrow-band interference 
 would be mistaken for an actual \hi\ feature.

\subsection{Selection of Directions}

     Long integrations in search of faint high-velocity 
clouds can be done in any direction, though we wished to concentrate on
 moderate and high Galactic latitudes where the high-velocity clouds
are most clearly distinguished from phenomena in the Galactic disk.  
One-third of the survey directions were toward 
extragalactic objects which are bright in the optical and UV, 
in the hopes that spectra at these
wavelengths might someday be obtained to complement the radio data.   
These directions are listed in Table 1.   Many bright QSOs and AGN have
been observed in \hi\ as part of another project 
(Murphy, Sembach, \& Lockman, in preparation) and were not repeated here. 
Only a few of the sources have significant radio continuum emission, 
and none were strong enough, at the 140 Foot Telescope, to cause
the  \hi\ to appear in absorption.

Each observation toward an extragalactic object was accompanied by
two flanking observations, of similar sensitivity, in directions offset 
 $\pm 6^m \ 52^s = \pm 1\fdg7 \ cos(\delta)$ in right ascension, 
making ``triples'' of measurements across the sky 
(some high-declination directions
are offset by twice this amount).  
The flanking directions were used as reference spectra for 
position-switched observations, they 
provide independent samples of the high-velocity sky, 
and they give some indication of the angular size of detected 
lines.  Directions are identified by the name of the extragalactic object,
 with the suffix ``+'' or ``$-$'' to show flanking positions.
A high-velocity line might be present in any or all of these 
spectra.  The observations were made as far south as declination
$-43\fdg3$; the median declination is $+17\arcdeg$.
We emphasize that the \hi\ lines detected in this survey have 
nothing to do with the background objects, most of which have a 
redshift well outside the velocity coverage of the survey.

\subsection{Observational Procedure}

  The observations consisted of a basic cycle in which a total power 
HI spectrum was measured for 6 minutes at each direction 
of a triple: the ``$-$'', source (S), and the ``$+$'' positions. 
This procedure was repeated at least six times for a total of no less than 
36 minutes integration at each location.  During data reduction, the 
total power spectra were combined to form position-switched pairs using 
each direction in a triple as either ``signal" or ``reference" with 
respect to the other two.   This procedure effectively cancels
all stray radiation in the spectra and much of the normal, low-velocity 
Galactic disk \hi, and usually gives excellent instrumental baselines. 
The 6 minute integration period, together with the $6^m \ 52^s$ 
separation of the positions in right ascension, meant that the telescope
tracked over the same azimuth and elevation for each position in a 
triple (when allowance for observing  overhead is included)
 helping to insure good instrumental baselines. 
Previous surveys have used special ``emission-free'' directions for
reference \citep{hw} or have modeled the bandpass \citep{barnes01}, 
but these techniques do not always give the very flat baselines 
needed to detect very weak lines.  
The disadvantage of using nearby reference positions, however, is that 
 a high-velocity cloud may be  extended over several degrees and 
thus cover all three locations of the triple.
These situations were resolved by making frequency-switched measurements,
or by choosing an alternate reference position as described below.
In all, total-power \hi\ spectra were measured in 860 directions, 
which sample a total  area of 60 square degrees.

About 100 directions were observed in frequency-switched as well as 
position-switched modes. 
Some of the frequency-switched spectra were taken concurrently with 
the position-switched data, and others were taken subsequently to restore 
emission which extended over an entire triple.  
 The frequency-switched spectra have  integration
times between 20 minutes and several hours.  Unlike the position-switched
spectra, the frequency-switched spectra can contain complete information on 
low-velocity \hi, but they also contain  stray radiation ($\S2.6$).

\subsection{ Data reduction}
     The spectra were calibrated by bootstrapping to standard 
regions S6 and S8,  and a
correction for atmospheric opacity at 21cm was applied 
\citep{williams}.  The frequency-switched
spectra were corrected for stray radiation (see $\S2.6$) using the method of 
\citet{kmr} as applied to the 140 Foot Telescope 
 by Murphy, Sembach, \& Lockman (in preparation).  A low-order
polynomial was fit to the spectra to establish the instrumental baseline.
 The partial cancellation of strong, low-velocity Galactic \hi\ 
emission in the position-switched spectra often produced ringing
 which was suppressed by smoothing with a Hanning function.

The position-switched spectra were examined for emission, and in 
cases where only one spectrum of a triple showed high-velocity
lines, the other two spectra were combined into a single reference
spectrum to reduce the noise.  In many cases where 
high-velocity gas was detected at all locations in a triple, a 
frequency-switched spectrum was obtained in one of the directions 
and used to resolve ambiguities and  restore the other two spectra.

Frequency-switched data were not available for all directions that had
extended high-velocity emission, so it was sometimes necessary 
 to rereduce the data using a reference spectrum taken many degrees away.
Because most directions observed in this survey 
did not have detectable  high-velocity
\hi, it was usually possible to find a set of suitable spectra 
taken on the same day for use as an alternate reference.  A spectrum
formed from an alternate reference, however, invariably has a poorer
instrumental baseline than one formed from the members of the same triple.

     \hi\ emission at $|V_{LSR}|>100$ km~s$^{-1}$ 
is found in two forms: spectral components well-separated from 
bright disk emission, and extended wings on the lower-velocity 
emission which cannot be described by a Gaussian function.  
We call these two cases  ``components"  and
``wings".  Occasionally a spectrum shows a clear \hi\ component 
whose  LSR velocity at the line peak is slightly less than the
100 km~s$^{-1}$ cutoff adopted as the definition of high-velocity gas. 
In this case we fit a Gaussian function to the line if it 
has significant emission at $|V_{LSR}|>100$ km~s$^{-1}$ and if it is likely
that the Gaussian parameters have physical meaning.  Wings are
characterized only by the first and zeroth moments of the velocity-weighted
brightness temperature.  Several lines were decomposed into multiple
components when the peaks of the components were clearly resolved,
if this procedure was reasonable given the noise in the data.  

The quality of the data is generally excellent.  
There were a handful of spectra
that showed some interference (see the example in \S 3.2), but it was 
usually so weak, narrow-band, and infrequent that it did not materially 
affect the measurements.  The worst problems came from occasional
bad instrumental baselines caused when observations were made too close to 
the Sun, and, in one entire observing session, 
by a problem in one half of the autocorrelator
which caused a kink in the baseline more than 100 km~s$^{-1}$ wide.  
In most cases no attempt was made to salvage the bad data, it was
simply removed from the survey with the consequence that the affected
directions had somewhat less sensitivity than the rest. A few
spectra show residual effects of this problem.

\subsection{Sensitivity and Detection Limits}
 The distribution of the rms noise in the emission-free baseline regions 
of the  final position-switched spectra,
 which include both smoothed and unsmoothed data, is shown in Figure 1.
  The median value of $\sigma_b$ is 3.4 mK.  
The long tail to high rms is caused by a 
few directions where radio continuum increased 
the total system temperature, a few directions with short integrations,
and  a few directions where the baseline
quality was not good because of equipment problems or solar interference.
Table 2 summarizes the parameters of the spectra, and gives the number 
of spectra in the final data set derived from each mode of observing. The 
frequency-switched spectra are not as homogeneous as the position-switched 
spectra because their  integration times vary considerably.
  The factor of two difference 
between $\sigma_b$ of the smoothed and unsmoothed frequency-switched 
data shown in Table 2 is a fluke caused by small numbers of smoothed 
spectra which included several very long integrations. 

     In order to determine our subjective ability to 
recognize weak lines in noisy spectra,  
simulated data were created containing Gaussian functions of random 
intensities at random velocities.  
These were  analyzed as if they were real data.  
This effort showed that lines with a  peak intensity 
$\geq 3 \sigma_b$  were nearly always detected, and at this level 
 there were virtually no noise features mistakenly identified 
as lines.  We therefore assume that
a line with T$_{pk}\geq 3 \sigma_b$ in any spectrum 
is real.  The lines detected in the survey under this criterion 
have  peak brightness temperatures between 8 and
2052 mK with a median of 57 mK.

Although the simulations suggest that all lines with 
T$_{pk} \geq 3\sigma_b$ will be detected, to estimate the
completeness of the survey we adopt a more conservative limit 
of $4 \sigma_b$, which is  $T_{pk} =  14$  mK
for the median value of $\sigma_b$. 
A line of this peak brightness temperature whose 
 width is 30 km~s$^{-1}$ has $N_{HI} = 8 \times 10^{17}$ cm$^{-2}$, 
and this value is therefore adopted as the median 
completeness level of the survey 
for emission which fills the antenna beam. A line at this brightness 
will be $\geq 3 \sigma_b$ for more than $70\%$ of the observations.

The survey is a factor of three to eight times more sensitive 
than that of Hulsbosch and Wakker (1988), 
 depending on whether the \hi\ cloud being observed is larger than 
$35'$  or smaller than $21'$, respectively.  The survey is a factor of 
1.1 to 2 times more sensitive to broad lines than the Southern 
survey of \citet{pg99} for clouds with a size $<15\arcmin$ and
$>21\arcmin$, respectively, and has a greater advantage for narrow
lines, because the Southern survey used a relatively coarse velocity
resolution of $\Delta v = 26.4$ km~s$^{-1}$, 
which will substantially dilute the peak signal in nearly half of 
the directions.

\subsection{Stray Radiation as a Source of Error}

Stray radiation is 21cm \hi\ emission which is received 
through an antenna sidelobe, rather than through the
main beam.  The sidelobe  may lie tens of
degrees away from the direction that the telescope is pointed.  
Stray radiation can be a significant component of high-latitude 21cm
spectra, and it often appears  as spurious wings on a profile 
which may extend more than  100 km~s$^{-1}$ from the line center 
(e.g., \citet{kmr,ljm}).  Figure 2 shows a 
140 Foot frequency-switched spectrum 
toward Ton 34 to illustrate the effect.  The 
curve drawn through the data is an estimate of the stray component 
calculated using the method described by Murphy, Sembach, \& Lockman 
(in preparation). In this 
 spectrum, there is $2.4 \times 10^{18}$ and $6.7 \times 10^{17}$
 cm$^{-2}$ of stray emission at $V_{LSR} \leq -100$ and 
$V_{LSR} \geq +100$ km~s$^{-1}$, respectively.

  Figure 3 shows the average amount of stray radiation 
that was removed from the $\sim 100$ frequency-switched observations in the
survey, expressed as an equivalent $N_{HI}$ integrated over 
$ |V_{LSR}| \geq V'$ for various values of $V'$. 
 The stray radiation at positive and negative velocities is 
distinguished by different symbols.  
In this sample, which should be typical of the northern sky, 
stray radiation is more significant at negative velocities because 
Galactic rotation produces a general preponderance of negative velocity \hi\  
 at positive declinations, 
some of which leaks in through sidelobes to produce the stray wings.
 For a similar telescope in the southern hemisphere the proportions 
will most likely be reversed.

 Stray radiation can mimic high-velocity \hi\   
at the level of a few $10^{18}$ cm$^{-2}$ 
(the stray component at lower velocities can be 
orders of magnitude larger).  The 
frequency-switched spectra in the survey were corrected 
by calculating the amount of stray radiation expected in 
each spectrum from a model
of the all-sky response of the 140 Foot Telescope convolved
 with all-sky surveys
of \hi\ (Murphy, Sembach, \& Lockman, in preparation).  
The calculated stray spectrum, which applies in
a particular direction at a particular time, is then subtracted from
the observed data.  This technique has some limitations, for 
the far sidelobe pattern of a large conventional radio telescope 
cannot be determined to high precision, but it does reduce 
the effects of stray radiation in  frequency-switched spectra by at least 
an order of magnitude (e.g., \citet{kmr}).

Sidelobes which are responsible
for  stray radiation are quite broad, so the stray component of \hi\ 
spectra taken closely in position and time will be quite similar, and will be 
eliminated in the usual position-switching difference technique.
But in directions where  high-velocity gas is  spatially extended, 
and the reference spectrum  must be taken some distance away,  cancellation 
of the stray radiation will not be complete.  This may result in 
a residual which is either positive or negative at a given velocity. 
In our data, faint wings on spectra reduced with  an alternate reference 
when there are no frequency-switched data  
should be viewed with caution.  These are identified 
by the note ``A'' in the data table.

\section{Results}

\subsection{The Data Table}

     Detections of high-velocity \hi\  are summarized in Table 3. 
 Directions are identified by the common name of the extragalactic
source upon which they were centered, with a ``$-$'' 
or ``+'' suffix to indicate flanking directions in a triple.  
Thus the directions labeled Q~022+0015$-$, Q~022+0015, and  Q~022+0015+ 
have $\alpha(B1950) =  00^h15^m18^s, 00^h22^m10^s$, and $00^h29^m02^s$, 
respectively, and all have essentially the same declination: 
$\delta(B1950) =  +00\arcdeg15\arcmin44\arcsec$. 
 The positions in the Table are those which were 
actually observed, and may differ from cataloged source
positions by small amounts.  Information in the Table is derived 
from unsmoothed, 
position-switched observations using a member of the same triple as the 
reference spectrum unless otherwise indicated.  About 15\%
of the entries come from frequency-switched observations.

     Line parameters were derived by fitting a Gaussian function. 
The line peak, $T_{pk}$, is a  beam-averaged brightness temperature.   
Peak temperatures and widths are
uncorrected for instrumental broadening, which can be 
significant for the narrowest lines (see $\S 3.6$).   The error estimates 
on the line parameters (given in parentheses) 
 are one standard deviation.  The quantity $\sigma_g$ 
is the rms  of the Gaussian fit with respect to the data, and $\sigma_b$ is the
rms noise in the part of the spectrum used to determine the 
instrumental baseline.  
The quantities $N_{HI}$ and $<V>$ are described in $\S3.1.1$. 
If a detected line could be associated with a known galaxy its name 
is given in the Table, and line parameters are listed without 
error estimates.  These objects are discussed in $\S3.5$. 

The notes to Table 3 indicate if an alternate reference position was
used for a position-switched spectrum (``A''), if the line parameters
were derived from frequency-switched (``F'') or Hanning-smoothed 
(``H'') data, or if the detection is uncertain (``U''). 
Many of the ``uncertain'' lines are wings near $\pm 100$ km s$^{-1}$ 
 on position-switched spectra that had to be reduced with 
an alternate reference position.

\subsubsection{Description of Wings and other Components}

Some spectra have emission 
at $|V_{LSR}|>100$ km~s$^{-1}$ which is blended with lower velocity emission
and cannot be described meaningfully by a Gaussian function. 
These blended lines will be called ``wings''.  
Table 3 gives the zeroth and first moments of the wings, 
 $N_{HI}$  and  $<V> \equiv \int N_{HI}(v)\ v \ dv / \int N_{HI}(v) dv$,
both calculated from $\pm 100$ km~s$^{-1}$ to a velocity where 
the emission becomes undetectable.   The quantity  $<V>$ 
contains information on the shape of the wing past $\pm100$ km~s$^{-1}$, 
and also, 
by its sign, shows the side of the profile on which the wing is found.  

The $N_{HI}$ and $<V>$  columns in Table 3 are also used occasionally 
 to describe  \hi\ emission 
in a spectrum which is not necessarily a wing, but 
cannot be modeled by a Gaussian function either.  
A specific example is discussed in $\S3.2$.  To distinguish this 
circumstance from a wing, the following convention has been adopted. 
If there is an entry for $N_{HI}$ in Table 3 and no Gaussian parameters
{\it in the same row}, this signifies that the entry is a wing
whose $N_{HI}$ has been counted separately from other components which 
may occur in that spectrum.  
For example, the spectrum labeled 4C 49.48+ described on rows 2 and
3 of Table 3 contains two spectral components:
a Gaussian (row 2) and a wing (row 3).  They are both at negative velocity.
The total $N_{HI}$ at $V_{LSR} < -100$ km~s$^{-1}$ in this direction is the 
sum of the $N_{HI}$ in the two entries, or $1.35 \times 10^{19}$ cm$^{-2}$.  
If, however,  
a value for $N_{HI}$ is given on a row which also contains information 
about a Gaussian component, then that value of $N_{HI}$ includes the emission 
in the Gaussian, and in any other Gaussian which has the same sign of velocity. 
This circumstance is rarer, but occurs, for example, toward
  Mrk $380-$ in rows 86 and 87 of Table 3.  Here the total $N_{HI}$ 
on the negative velocity side of the line is $4.87 \times 10^{19}$
 cm$^{-2}$,  which includes the two Gaussian components,
and additional emission not contained in either.  
The spectrum in this particular direction is illustrated 
in $\S 3.2$ and discussed further there.

\subsubsection{Table of Gaussian Components}

The lines detected
in this survey that are well described by a Gaussian
function are given in Table 4.  The \hi\ column density $N_{HI}$ was
calculated under the optically thin assumption: 
$ N_{HI} = 1.823 \times 10^{18} \ T_{pk} \  1.065 \  \Delta v$. 
It includes all emission in the component, even that which 
lies  at $|V_{LSR}| < 100$ km~s$^{-1}$.  
Table 4 contains no line wings or other components 
which could not be represented by a Gaussian, and it  does not 
contain lines that are associated with known  galaxies or galaxy groups.

\subsection{Sample Spectra}

  The spectra shown in Figures 4, 5,  and 6  
illustrate the types of \hi\ lines 
that have been detected.  Only the central velocities
 of the spectra are shown. 

Figure 4 shows position-switched data containing some of the weaker 
lines detected in the survey.  The lower two spectra have
been Hanning smoothed. 
The spectrum in the upper panel contains two blended 
high-velocity \hi\ lines near $-300$ km~s$^{-1}$ and the 
Gaussian functions that were fit to them.    The lines 
have $N_{HI} = 8.3 \times 10^{17}$  and 
$1.5 \times 10^{18}$ cm$^{-2}$.  Position-switching to a nearby reference has 
canceled most emission at  $|V_{LSR}|<100$ km~s$^{-1}$.

	The spectrum in the central panel of Fig. 4  has 
emission from a high-velocity cloud at $V_{LSR} \approx -400$ km~s$^{-1}$,
together with a double-peaked \hi\  line from the galaxy UGC 64 
 which serendipitously lies in this direction at $+300$ km~s$^{-1}$.  
Low-velocity \hi\ emission is confused because of 
cancellation in the position-switching, but the negative signal near 
$V_{LSR} \sim -100$ km~s$^{-1}$ is actually an
emission wing in the reference spectrum (4C 40.01+), with a total 
$N_{HI} = 2.5 \times 10^{18}$ between $-150$ and $-100$ km~s$^{-1}$ 
(see row 10 of Table 3). The lower panel of Figure 4 shows a spectrum 
with  a high-velocity \hi\ line at $-327$ km~s$^{-1}$ that has $\Delta v = 30$
km~s$^{-1}$, the median width of the lines detected in this survey. 

Figure 5 shows two Hanning smoothed, position-switched 
spectra from the survey which illustrate
profile wings and emission which cannot be described by a
Gaussian function.  The spectrum in the upper panel  has  a line at 
+145 km~s$^{-1}$ with $N_{HI} = 1.25 \times 10^{18}$ cm$^{-2}$.  
It has  $\Delta v = 23$ km~s$^{-1}$, and is 
 well separated from the rest of the emission.  This spectrum
also contains a second component whose peak is near +85 km~s$^{-1}$ 
with an extension past +100 km~s$^{-1}$.  This line may have
been partly canceled in the position-switching, so its true shape
is uncertain, and it is listed in Table 3 as 
a wing.  The total $N_{HI}$ at $V_{LSR} > 100$ km~s$^{-1}$ in this direction 
is the sum of the wing (Table 3 row 205) and the 
Gaussian component  (row 206).  This spectrum also shows some 
residual instrumental baseline structure at positive velocities.

The lower panel of Figure 5 shows a more complex spectrum.
 The  high-velocity \hi\ in this part of the sky extends 
over many degrees, so reference spectra were chosen from a triple that
 was  about 12 degrees distant. The spectrum is thus noted ``A'' in 
Table 3.   The two Gaussians fit to the data give a reasonably
accurate representation of 
much, but by no means all, of the high-velocity emission.
  While a third Gaussian component could 
have been fit near $-240$ km~s$^{-1}$, 
its properties would not have been well constrained, so  
 a total integral for all the emission at $V_{LSR} < -100$ km~s$^{-1}$ 
was calculated instead and is given in row 86 of Table 3.  In the
convention adopted for Table 3, whenever a value for $N_{HI}$ is 
listed in the same row as a Gaussian component, that component, and
all others on the same side of the profile, are included in the 
value of $N_{HI}$.  Thus the total $N_{HI}$ at negative velocities
toward Mrk $180-$ is $4.87 \times 10^{19}$ cm$^{-2}$,  comprised
of two Gaussians and the additional emission (see $\S 3.1.1$).

Figure 6 shows representative frequency-switched spectra  
after their correction for stray radiation.  
The top spectrum is unsmoothed at a channel width of 1.25 km~s$^{-1}$ and
has $\sigma_b = 42$ mK.  The component shown with the Gaussian fit 
is listed in Table 3 even though its velocity at line peak is only 
84 km~s$^{-1}$,  because much  of it lies at 
$V_{LSR} < -100 $ km~s$^{-1}$ and it is well separated from normal 
Galactic disk \hi. 

Spectra in the lower three panels of Fig. 6 were taken $\pm0\fdg8$ 
on either side of, and toward, the galaxy Markarian 205.  The 
high-velocity line toward Mrk $205-$  
has a  width $\Delta v = 14.6$ km~s$^{-1}$ and the spectrum 
shows a rare example of radio interference  near $+40$ 
km~s$^{-1}$.  The  high-velocity \hi\  seen directly 
toward Mrk 205 has been decomposed into two Gaussians with $\Delta v = 
7.1$ and $18.0$ km~s$^{-1}$.  This direction provides an example of the 
multiple component (core-halo) structures found in some high-velocity spectra 
\citep{cram76}.   The spectrum toward Mrk 205+,
 the only one in this Figure which is Hanning smoothed, 
contains a weak line with $\Delta v = 4$ km~s$^{-1}$, 
the narrowest in this survey. 
Even narrower lines have been observed in this particular 
high-velocity cloud at an angular
resolution of one arcmin \citep{bb00}.  
The emission feature at $-140$ km~s$^{-1}$ 
in the Mrk 205+ spectrum may be an artifact of the stray-radiation removal 
process as it could not be detected in the more sensitive position-switched
data.   There are several features  in other frequency-switched
spectra that may also be artifacts of stray radiation.  They are not
included in the data tables.

\subsection{Observed Directions}

Although some directions  as far south as $\delta = -43\fdg3$ have 
been observed, southern declinations are undersampled
compared to northern declinations.   The distribution of the 
observations in Galactic latitude is shown in Figure 7, with a curve 
 proportional to the amount of sky at each latitude with 
$\delta \geq -30\arcdeg$.  The curve is a reasonable approximation to the
distribution except at $|b|<20\arcdeg$, where there  are fewer 
observations than would be found in a uniform sample.  This 
is a consequence of our observation toward  UV and optically 
selected extragalactic targets, which are unlikely to be found in 
 directions of high extinction and thus avoid the Galactic plane.

\subsection{Notes on Specific Spectra}

Table 3 includes lines in three directions (rows 58, 377, and 378) 
which have $T_{pk} < 3 \sigma_b$ and therefore fall below our
limit for a claimed detection.  They all have $T_{pk} \approx 7$ mK and
values of $N_{HI}$ in the range $4 - 6 \times 10^{17}$ cm$^{-2}$, among the
lowest in the survey.  Although these lines are not used in any
analysis, they appear to be real and have been included in the Table.

Table 3 also contains three lines with $T_{pk} > 3 \sigma_b$ which
we suspect are spurious, likely to be an instrumental artifact
of some undetermined sort, even though they are above our detection limit.
These have the smallest values of $N_{HI}$ in the survey.  Two of the 
lines are toward Mrk 992 (rows 38 and 39).  With 
LSR velocities of $-492$ and +465 km~s$^{-1}$, these have the second most
negative, and the most positive,  velocity of 
any line in the entire survey.  The 
positive velocity line lies almost 200 km~s$^{-1}$ 
beyond the next-highest velocity line, 
 and it sits in an area of 
longitude-velocity space occupied only by galaxies that are
part of  the Hubble flow (see \S 4).  
It strikes us as improbable 
that one spectrum should contain two such extreme lines, 
and such weak lines, though scrutiny of the data gives us
 no reason to reject them outright. 

The line toward PKS $0859-14+$ (row 128) has V$_{LSR} = -696$ km~s$^{-1}$.  
This  is almost 200 km~s$^{-1}$ more negative than our 
secure detections, and, 
at $T_{pk} = 10.3$ mK and $N_{HI} = 3.8 \times 10^{17}$, the line is among the 
weakest in the survey.  
Moreover, most high-velocity \hi\  at its longitude and
latitude  has positive LSR velocity, not negative. 
Examination of the raw data, however, 
 give us no reason to reject this line and it has
been kept in the survey. 

The nominal limit for detection of a line,  $T_{pk} > 3 \sigma_b$,
 was relaxed for one line in the group from 4C50.$43-$ 
which has $T_{pk} = 2.8 \sigma_b = 3.1 \sigma_g$.  
This direction was observed using frequency-switching,  
the line appears in many channels,  and it is most likely real.

\subsection{Galaxies}

	In at least twelve  directions 
\hi\ lines were detected which can be 
attributed to a known galaxy in the antenna beam (this does not 
count the Magellanic Stream, which is discussed in $\S 6.1$).  
The galaxy M31 was detected in two separate spectra.  
An example of \hi\ emission from a galaxy is shown in Figure 4.
In Table 3,  galaxies are identified by a note and their 
peak line temperature and velocity  are estimated directly from the 
data,  not from a Gaussian fit. For galaxies, the approximate total velocity
extent is given in place of $\Delta v$, and, to avoid any possible 
confusion with high-velocity clouds, the errors on the measured
quantities are set to zero.  Lines from galaxies 
are not included in any analysis of the survey results.

It is also likely that gas in the M81/M82 system was detected in the survey. 
 The three  directions associated with  $0959+68$W1 lie along the southern 
edge of the M81/M82 group, which contains several gas-rich galaxies as 
well as \hi\ distributed throughout the group, probably the result of a 
tidal interaction \citep{apple,yun}.  We detect \hi\ in 
 the direction of 0959+68W1+ at the edge of the galaxy group, implying 
 that gas in the group is more extensive than was previously known 
at the level of a few $10^{18}$ cm$^{-2}$.

\subsection{Line Width}

 The distribution of measured line widths is shown in Figure 8, 
after correction for the channel width of the spectrometer.  
The median corrected value,  $\Delta v = 30.3$ km~s$^{-1}$, is marked
with an arrow.  There is no correlation between $\Delta v$ and 
the quantities $V_{LSR}$, Galactic latitude, or signal-to-noise
ratio of the line.  There is a tendency for the broadest lines to have a 
lower peak antenna temperature than the average.
Some of the the broadest  lines are probably blends of separate emission
 clouds in the antenna beam (see also \citet{mirabel81, bb00}).  
Many of the narrow lines would have been substantially diluted in some 
 previous surveys made at a velocity resolution of 
$15 - 20$ km~s$^{-1}$, leading to a significant undercounting
 of this component of high-velocity gas. 

 Only 1\% of the lines detected here have a width  
$\leq10$ km~s$^{-1}$.  \citet{colgan}, 
 in observations with a $4\arcmin$ beam, 
found  that 6\% of their lines were this narrow,
and that they originated in localized regions of median size 
$\sim 10\arcmin$.  This suggests that narrow lines may be 
underrepresented even in the current observations because of beam smearing.

\section{The Sky Coverage of High-velocity \hi}

\subsection{All High-Velocity \hi}

The goal of this survey was to determine the fraction of the sky which
is covered by high-velocity \hi.
Figure 9 shows an Aitoff projection  in 
Galactic coordinates, with solid symbols marking directions with
detected high-velocity gas, either  
Gaussian components or wings, and open circles
marking directions with no detection.  The organization of the observations
into ``triples'' is apparent here, as is the zone of avoidance along the 
Galactic equator and the region at high longitudes which is 
below the horizon at Green Bank.  Figure 10 
shows the detections divided into wings and Gaussians 
components, where the wings 
are emission lines blended with lower-velocity gas.  Some of the wings
lie at distinctly low latitude and may be extensions of the Galactic disk
rather than emission from discrete clouds.

Figure 11 shows the percent of the sky covered by high-velocity \hi\ 
 as a function of log($N_{HI}$).  For this accounting, the total amount of
\hi\ at $|V_{LSR}|>100$ km~s$^{-1}$ in each spectrum was summed 
over both positive and negative velocity, and the normalization is
 the total number of directions observed.  The arrow marks
the adopted median 
completeness limit of the survey, $8 \times 10^{17}$ cm$^{-2}$.
For comparison, the dashed lines show the percentage sky coverage 
from the \citet{hw} survey, derived from the statistics on 
peak line temperatures of \citet{wakk91}, under the assumption that 
 the lines in that survey had $\Delta v = 30.3$ km~s$^{-1}$, the median
from our survey. The nominal completeness level of the 
\citet{hw} survey is  $2 \times 10^{18}$ cm$^{-2}$, shown 
by a dashed arrow.
The Wakker compilation does not include any emission from 
the Magellanic Stream, which, from our data, 
would add $\sim4\%$ to the total area under the dashed curve.

There is general agreement between the Hulsbosch-Wakker data and the current
survey for $N_{HI} \geq 6 \times 10^{18}$ cm$^{-2}$, but 
below this level we have found 
 many \hi\ lines that are too weak to have been detected before.  
  Our data are consistent with
a constant number of high-velocity \hi\ lines, per logarithmic interval
in $N_{HI}$, from $10^{18.6}$ cm$^{-2}$ down to the 
sensitivity limit of the survey, and the sudden decrease in 
the number of lines at that limit  suggests that there
may be many high-velocity \hi\ clouds at even lower $N_{HI}$.

Of the 860 directions observed,  37\% (316),  show some high-velocity \hi. 
 There are 242 directions, or 28\% of the total, that have high-velocity
\hi\ which can be characterized by a Gaussian component well-separated from 
the disk emission.  
If the abundance of high-velocity lines in the decade of $N_{HI}$
below the completeness limit is similar to that in the decade above
the limit, it would imply that about $60\%$ of all sight-lines on 
the sky will intersect a 
 high-velocity \hi\ cloud with  $N_{HI} \geq 10^{17}$ cm$^{-2}$.

\subsection{The Gaussian Components}

The 309 high-velocity \hi\ lines in the survey that were fit with 
 Gaussian functions come from 242 directions.  The distribution of their 
column densities (from Table 4) is shown in Figure 12 with a solid curve 
for all the data and a dashed curve for only the higher velocity lines.
To first order, the curves are simply scaled copies of each other, 
suggesting that there is no strong relationship between the velocity
of lines and their $N_{HI}$.  
This topic is discussed further in a following section.
  Table 5 summarizes the measured values of $N_{HI}$ for the components.  
A given direction will contribute  more than once to Table 5 if
it contains multiple lines  ($\S5.3$).  The median $N_{HI}$ of the 
Gaussian high-velocity lines is $3.5 \times 10^{18}$ cm$^{-2}$.

\section{Kinematics of High-velocity Gas}

\subsection{Kinematics of the Components and Wings}

Figure 13 shows the velocity  of the lines, wings and
galaxies as a function of longitude.  The upper panel shows $V_{LSR}$, and 
the lower panel shows 
velocities with respect to the Galactic center, where 
$V_{GCSR} \equiv  V_{LSR}+220.0 \ sin(l) \ cos(b)$.  
In general, the galaxies (marked with open stars)
are part of the Hubble flow and occupy a different place in this diagram
than most high-velocity clouds.  A significant part of the LSR 
velocity of high-velocity gas arises in the rotation of the Galaxy, as 
shown by the more even distribution of points about zero $V_{GCSR}$ 
than about zero $V_{LSR}$.  The appropriate velocity reference 
frame for high-velocity \hi\ will be considered further in the 
next section. 

Figure 14 shows the LSR velocity of the galaxies and the Gaussian 
high-velocity components from Table 4  vs. their total hydrogen column density.
The  weakest lines have velocities  similar to the stronger lines, 
though no line with $N_{HI} >10^{19.5}$ cm$^{-2}$ has a very high velocity.
Not too much weight should be placed on the 
three points with the lowest $N_{HI}$, as their accuracy  is 
suspect ($\S 3.4$).  Previous surveys have found that 
high-velocity \hi\ is confined 
within $-465 \leq V_{LSR} \leq +420$ km~s$^{-1}$  
\citep{wakk91, morras2000}.  The current detections, with the exception of a 
few suspect lines, lie within the same range, indicating that the
newly-discovered faint high-velocity \hi\ lines are not a kinematically
distinct population, but have the same general
velocity bounds as the brighter lines.

Negative velocities dominate our detections, 
in part because of selection effects 
 -- many known high-velocity clouds with positive velocity 
are at very negative declinations below the survey limit -- 
but also because there are a number of lines with 
large negative velocity around longitude $100\arcdeg$.  
Many of these arise in the Magellanic Stream.

\subsection{Kinematics of the Integrated Emission}

The upper panel of Figure 15 shows a composite spectrum made by summing 
all the emission in the  Gaussian components from Table 4.
 The vertical dashed lines mark the limit of the definition of 
high-velocity gas in the LSR frame.  
This Figure does not include the emission
 in wings,  so it is  incomplete near $\pm100$ km~s$^{-1}$, 
but apart from this, it is directly proportional to 
 the likelihood of observing high-velocity \hi\ 
 at a given LSR velocity in the northern sky as determined by 
 the survey data.   The cross shows the
mean velocity of the emission and $\pm 1 \sigma$.  In these data 
negative LSR velocities are dominant and the emission extends to a much
greater negative than positive LSR velocity.  The two lower panels
show the emission summed in the Galactic Center Standard of Rest 
 (defined in the previous section) and in 
the Local Group Standard of Rest:
 $V_{LGSR} \equiv V_{GCSR} +40sin(\ell)cos(b)
-62cos(\ell)cos(b)$ km~s$^{-1}$ \citep{bb99}.

High-velocity \hi\ emission in the northern sky is 
predominantly at negative LSR velocities, but 
when the components are averaged with respect to the Galactic Center, 
 they are remarkably symmetric, with a 
mean velocity near zero and a substantially reduced dispersion.  The
extreme velocities have nearly identical magnitude.  When the emission
is further reduced to the Local Group Standard of Rest it has a somewhat
broader extent and is skewed to positive velocity. 
Table 6 gives the  mean and standard deviation of the composite 
spectra and  also  the mean and standard deviation of the 
\hi\ line velocities  with no weighting by the amount of \hi\ 
in the line.  Unlike the total emission, the velocities 
of the line peaks have a negative mean in all three reference systems.

It has been known for some time that the 
Galactic Center standard of rest is more appropriate for high-velocity
\hi\ than the LSR \citep{giovanelli, wakk91}.  
There have been recent suggestions that the Local Group Standard of Rest 
is even more appropriate, in that the mean velocity of a selected 
group of high-velocity clouds is nearer zero, and their dispersion about 
the mean smaller, in that system \citep{blitz, bruns, bb00}. 
The integrated emission, which we believe is a more robust indicator 
than counts of lines or clouds,  does not support the Local Group 
Standard, and indicates that the GCSR system is preferred (see also
the analysis of a different population in \citet{putmanetal2002}). 
In the GCSR system, the integrated emission has a mean velocity near zero and
a dispersion smaller than in the other two systems.

\subsection{Directions with Multiple Components}

Some of the sample 
spectra shown in Figures 4, 5, and 6 contain several high-velocity 
\hi\ lines.  In the entire survey, 180 directions have a single line, 
57 (or 24\%) have two, and 5 have three, not counting wings.  
The separation of these  components in 
velocity is shown in Figure 16.  The median separation is 
48 km~s$^{-1}$, similar to the separation of the two lines in the 
upper panel of Fig. 4.  
Given that the median value of the linewidth is 30 km~s$^{-1}$,
many of the blends would not be resolved at a velocity resolution 
much coarser than that used here.  The location of directions with 
multiple lines is shown in Figure 17.  Most multiple lines do not 
seem to be chance superpositions of unrelated clouds, but 
are found toward  high-velocity complexes such as
Complex A and the Magellanic Stream, and reflect their internal
velocity structure (e.g., \citet{mirabel81, wayte, murphy00, 
richteretal2001}).

The \hi\ lines from some high-velocity clouds have two components
at a similar  velocity, one broad and one narrow 
 \citep{cram76, bb00}.  This has been interpreted as evidence for 
a core-halo thermal structure in the 
clouds \citep{ferrara94,wvw97}.
Lines like this are rare in the survey.  They would be listed as multiple
components with a very small velocity separation;  as Fig.~16 shows,  
this is seen in  only a few percent of the spectra. 
The best core-halo example is Mrk 205 (Fig.~6), which has been studied
at high angular resolution by \citet{bb00}.

\section{Association with High-Velocity Cloud Complexes}

The high-velocity \hi\ sky is dominated by large complexes that 
have ordered velocities over 
large areas  \citep{giovanelli, wvw91}.
Some complexes, like the Magellanic Stream (MS) and Complex C, 
 are certainly physical associations with a distinct pattern of abundances 
and likely a single origin (e.g., \citet{richteretal2001}).  Others, like the 
 Outer Arm (OA) complex, may be part of the Galactic warp 
\citep{habing, davies, verschuur75, haud}.  
Some high-velocity clouds have  been grouped into broadly-defined 
``populations'' such as  those at 
``extreme-negative'' (EN) or extreme positive 
velocities \citep{wvw91}, which may be merely collections of 
unrelated objects.  

All of the Gaussian components of Table 4 have been examined for
their possible association with a known high-velocity \hi\ complex 
using the definitions of \citet{wvw91} as a guide, 
 and the results are given in the final
column.    There is some ambiguity in the identifications 
because the boundaries of the complexes are not sharp, and are not well 
known at the sensitivity limits of the survey. 
Lines from a ``population'' are identified (parenthetically), 
but for statistical purposes are not included with lines from the complexes.
The upper panel of 
Figure 18 shows  the high-velocity \hi\ lines
detected in this survey coded by velocity interval, illustrating their 
 association with large  complexes. There is a high probability of membership
in a complex for 232 of the 309 lines, or 75\%, while for another 47 
lines (15\%)  an association is possible.
 
 The lower panel of Figure 18 
shows lines detected in the survey which could not be assigned to a complex, 
and Figure 19 shows V$_{LSR}$ vs. $N_{HI}$ for the  lines from 
Table 4, where open symbols are used to mark 
 lines not obviously associated with a known high-velocity \hi\ complex. 
The vast majority of lines at all values of $N_{HI}$ are associated with a 
complex.  The lines not in complexes do not have peculiar 
kinematics, and are distinguished mainly in that they have lower 
$N_{HI}$ than the average.  There are indications in our data 
that many complexes are more extensive than previously suspected.  Several 
specific examples are discussed below.

\subsection{An Extension of the Magellanic Stream}

The dominant  feature of the southern high-velocity \hi\ sky
is the Magellanic Stream.  It 
has been studied by Mathewson and many others 
\citep{mathewson, haynes79, mirabel81, morras, wayte, putman00}.  
The Magellanic Stream is a coherent 
structure extending  more than $100\arcdeg$
across the sky from the Magellanic clouds.  
Figure 20 shows the survey data for $b \leq 0\arcdeg$ with emphasis on the
Stream.   It has been proposed that nearly all of the 
very negative-velocity emission below the Galactic plane at these longitudes
comes from  the Stream \citep{mirabel81}, although we have not 
made this assignment in Table 4.  Filled stars show lines   that 
  are noted ``MS*'' in Table 4. These  lie beyond the traditional 
borders of the 
Stream but  are most probably an extension of it, in view of the location 
and kinematics of the lines and the recent theoretical work on
the expected extent of the Stream (\citet{gardiner96, gardiner99} and
references therein). 

A number of the lines which we believe are part of 
the Stream lie on the same great circle as the main feature, but 
about $10\arcdeg$ past its traditional tip near $b = -40\arcdeg$ 
\citep{wayte}.  The lines typically have an $N_{HI}$  of a 
few $10^{18}$ cm$^{-2}$.  
 Other filled stars in Fig.~20 mark a group of lines  possibly part of
the Stream at a similar latitude but higher longitude than the
main section.  They have the very high negative velocities characteristic
of the Stream, and theoretical 
 models suggest that the Stream may be distended in this direction 
\citep{gardiner99}.  Lines which are suggestive of the Stream because
of their location and 
high negative velocity are marked ``MS?'' in Table 4 and shown 
as open stars in Fig. 20.  

There are directions near the MS where we find  no high-velocity  
 \hi\ emission at all (Fig.~9), but there is other
evidence which suggests that the Stream may have considerable
extent at levels below our sensitivity limit.  
Toward PKS 0003+15 at $\ell,b = 107\fdg3-45\fdg3$ (entry 128 in Table 4)
we find a line with the velocity of the MS, but somewhat 
outside its traditional borders. About $5\arcdeg$ away, 
\citet{gibson2000} have detected Mg II in absorption at a similar 
velocity toward the galaxy III Zw 2, also in our survey, but 
that direction does not have  high-velocity \hi\ emission to a 
beam-averaged $3\sigma$ limit of $N_{HI} = 4.5 \times 10^{17}$  
cm$^{-2}$ for an assumed line width of 30 km~s$^{-1}$.
This implies an abundance ratio 
$N_{Mg II} / N_{HI} \leq 2 \times 10^{-5}$, and 
[Mg II/\hi]~$\geq -0.22$ for the Magellanic Stream in this direction, 
using solar abundances from \citet{ag89}. This abundance is 
consistent with that of the LMC and marginally consistent 
with the SMC, but the need for a 
 correction for ionization or dust depletion, the possibility of 
substantial angular structure in the \hi, and the fact that 
the \hi\ column is only an upper limit, make interpretation of this
ratio problematic (see \citet{gibson2000}).  Higher angular
resolution 21cm \hi\ observations toward III Zw 2 are needed 
to resolve this issue, but in any case, the Magellanic 
Stream probably covers considerably more area than even our 
measurements suggest.

 A leading arm stretching out from 
the Magellanic Clouds opposite the Stream has been discovered 
\citep{putman98,putman00}.  Its extent is not yet certain, but 
there have been suggestions that it encompasses 
 most  of the positive-velocity
emission in complex WD and population EP 
because of the kinematics and metallicity of the gas 
\citep{mathewson, west, Lu94, Lu98,sembach01}.	

\subsection{An Extension of Wright's Cloud near M33}

Wright's cloud was discovered in \hi\ emission close to 
galaxy M33.  It covers more than $5\arcdeg$ in right ascension at 
 velocities near $-400$ km~s$^{-1}$ 
(\citet{wright74, wright79}, see also Fig.~5 of \citet{blitz}).  
This object has been detected in the survey (noted `Wr' in 
Table 4)  up to  $3\fdg5$ beyond the known boundaries
of the cloud, suggesting that it is extended 
to lower right ascension (and lower Galactic longitude) 
at levels $N_{HI} \leq 2 \times 10^{18}$ cm$^{-2}$.  
Some of the weak lines have 
the same multi-component structure found in the strong lines, and 
their velocity varies almost linearly with position, merging smoothly with
the velocity of the main cloud.  Thus Wright's cloud, like the Magellanic
Stream, has a significantly increased extent at low column densities.  
With the new data, the linear size of this cloud 
 is $ > 0.15 \ d $ kpc, where $d$ is its distance
in kpc.  If it is at the distance of M33, it must be more than 
100 kpc across.  

Wright's cloud has a velocity similar to parts of the Magellanic Stream 
$\sim 25\arcdeg$ away, but differs in velocity from 
 the extension of the Stream 
which lies closer and at a more similar Galactic latitude.  

\subsection{Complex L}

Complex L is a small group of clouds lying in a line 
near $\ell,b = 347\arcdeg+34\arcdeg$ \citep{wvw91}.  We have detected \hi\ 
with a similar latitude and $V_{LSR}$ at several positions about 
$10\arcdeg$ away in longitude. 
This emission is quite distinct from the other \hi\ in the area.  The lines 
all have $N_{HI} \leq 4 \times 10^{18}$ cm$^{-2}$ and 
 are listed as ``L?'' in Table 4.

\subsection{The Edges of the Complexes} 

 Figure 21 shows the 
distribution on the sky of the Gaussian high-velocity lines divided 
 at the median $N_{HI} =  3.5 \times 10^{18}$ cm$^{-2}$ into 
high-$N_{HI}$ and low-$N_{HI}$ groups. No distinction is made for 
the velocity of the lines.   In many 
instances the low-$N_{HI}$ lines cluster around their brighter 
counterparts suggesting that they trace the indistinct edges  
of the dense concentrations.  This is particularly true for
the Magellanic Stream ($60\arcdeg \lesssim \ell \lesssim 120\arcdeg; 
b\lesssim -30\arcdeg$), but is also seen in the W Complexes, and Complexes C, 
A, and M as well.  One conspicuous exception is the Outer Arm (OA)
 Complex where $95\%$ of the lines are quite bright at 
$N_{HI} > 10^{19}$  cm$^{-2}$, and no lines have $N_{HI} < 
2 \times  10^{18}$  cm$^{-2}$.  This lends weight
to considerations that the OA Complex is not properly part of the
 high-velocity cloud phenomenon, but is somehow part of the 
warped Galactic disk (e.g., \citet{habing, haud}).

\section{Angular Structure of the Emission}

\subsection{Nature of the low-$N_{HI}$ Lines}

Do the weakest lines in the survey result from beam
dilution of small, bright \hi\ clouds, or do they 
actually arise in broadly-distributed low-$N_{HI}$ gas?  
A definitive answer cannot
be supplied from the survey data alone, but there are suggestions
that both circumstances are being observed.  Table 7 lists some 
representative directions where a high-velocity \hi\ line has been
detected in all three position of a triple, and at least one direction
has $N_{HI} \leq 2.0 \times 10^{18}$ cm$^{-2}$.  The angular separation 
 between each of the measurements is given 
in the last column.  The first
four entries in the Table show strong gradients in the observed 
$N_{HI}$,  suggesting that the lowest values might arise from 
beam smoothing, e.g., at the sharp edge of a bright cloud.  
This is already known to the case for Mrk 205, where  high
angular-resolution measurements  show \hi\ concentrations of 
$10^{21}$ cm$^{-2}$ with sizes of a few arc-min \citep{bb00}.  

The final four entries in Table 7 show more modest variations in $N_{HI}$ 
with angle, and there are directions, e.g., toward B2 0051+29, which give 
the impression of an  extended region of low
column density.  This behavior is the rule rather than the exception in our
survey.  If a line is seen at two adjacent positions in a triple and
one has $N_{HI} \leq 2 \times 10^{18}$ cm$^{-2}$, then the majority of the
time its neighbor will have $N_{HI} \leq 4 \times 10^{18}$ cm$^{-2}$.
Directions with low $N_{HI}$ are most often found near other directions 
with low $N_{HI}$.  The examples of extended low-$N_{HI}$ emission 
come from several complexes indicating that the phenomenon is not
an isolated one. 

\subsection{Angular Extent of the Emission}

The data in each triple can be used to make generalizations about the
angular extent of the objects detected in the survey.  For each
Gaussian in Table 4, we examined adjacent directions to see if 
emission was present a similar velocity.  After eliminating
lines which might be confused with nearby wings,
 the survey gives some information on angular size for 163 emission
lines.  The results are summarized in Table 8, which 
shows how often high-velocity emission is extended over several 
 positions in a triple.  
About 40\% of the lines are seen at more than
one position, and only 7\%  arise from objects which appear
isolated in the survey data, in that they are found 
only at the center position
of a triple and not at the edges.  These are candidate compact
high-velocity clouds \citep{bb99} and are listed in Table 9.
  Most are associated with complexes and thus are probably 
 knots in larger clouds, and not isolated,  compact objects.

Lines seen only at the edge  of a triple, i.e.,
  at either the `+' or `$-$' positions, 
are three times more frequent than lines seen only at the 
center position.  This suggests that the vast majority of
the lines detected in this survey come from objects whose extent is
more than a few degrees.  This is consistent with the results of 
\citet{colgan}, who measured a number of high-velocity clouds and 
found a median  size $\sim 2\arcdeg$  down to a few $10^{18}$ cm$^{-2}$.

\section{Mass}

The mass of high-velocity gas is uncertain because
of our ignorance of the distance to most of the complexes, 
but it is possible to estimate the relative 
importance of the faintest lines to the total.  The mass of an \hi\ 
cloud has the proportionality $M \propto N_{HI} \Omega d^2$ where 
$\Omega$ is the solid angle and $d$ is the distance.  The high degree
of association of the low $N_{HI}$ lines with the giant complexes, 
both in kinematics and spatially, suggests that there is no relationship 
in general between $N_{HI}$ and $d$, i.e., that the faint lines are 
located at the same distance as the bright lines which define the 
traditional complexes.  The survey data do not give reliable
estimates of $\Omega$, but we will assume that the angular size is
proportional to the frequency of occurrence of lines in the survey.  The
incremental \hi\ mass in a range of column density is thus
proportional to the total number of lines detected in that range times
the average column density: $M_i \propto n \langle N_{HI} \rangle_i = 
\Sigma N_{HI_i}$.  This quantity is given in Column 4 of Table 5, 
and is expressed as a percentage of the total $N_{HI}$ in Column 5.
Each interval of $N_{HI}$ above $10^{19}$ cm$^{-2}$ makes a significant 
contribution to the total (see also \citet{putman00a}), but below
this, the number of lines does not increase as fast as 
$N_{HI}^{-1}$, and their contribution to the total mass becomes insignificant.

Figure 22 shows the cumulative distribution of the numbers of 
detected lines (dashed curve) and the total $N_{HI}$ that they contain 
(solid curve) above various values of log($N_{HI}$).  
About 80\% of the total column density is contained in lines with
$N_{HI} > 10^{19}$ cm$^{-2}$ and only 10\% in lines with $N_{HI} \leq 5 \times 
10^{18}$ cm$^{-2}$, even though most of the detected lines lie at 
the lower column densities.  The same general results are found 
if specific complexes, like the OA or MS, are excluded from the data.
We conclude that the vast majority of faint lines 
does not contribute  significantly to the total mass of high-velocity \hi. 
Of course, this does not exclude the possibility that there are some 
 high-velocity clouds which are not 
associated with the complexes and lie at a much greater distance 
 thus containing significant mass (e.g., \citet{bb99, sembach99, bbc01}).

\section{Summary Comments}

The high-velocity \hi\ emission detected in this survey covers 37\% 
of the sky, and lines are found down to  the median completeness 
limit of $8 \times 10^{17}$ cm$^{-2}$ and below.  
There is a nearly constant covering fraction per unit log($N_{HI}$) from 
$N_{HI} = 10^{19}$ cm$^{-2}$ down to the limit of the survey,
and there is no indication  of a cutoff in the 
high-velocity \hi\ population at low $N_{HI}$. 
Extrapolating this trend  to still lower $N_{HI}$ implies  a covering fraction 
near $60\%$ for $N_{HI} \geq 10^{17}$ cm$^{-2}$.  It is clear that 
the limiting $N_{HI}$ of the high-velocity cloud phenomenon has yet 
to be established. 

High-velocity \hi\ emission lines can be characterized as either 
    ``wings'' or ``Gaussian 
components'', the difference being whether they are blended
with lower-velocity emission and thus have uncertain properties (the wings), 
or can be fit by a  Gaussian function whose parameters are likely to
be meaningful (the components).  There are 309 detected Gaussian 
components, found in 28\% of the survey directions.  
Much of our analysis concentrates 
 on the components (Table 4) because they have accurate velocities and their 
values of  $N_{HI}$ do not depend critically on the precise velocity 
definition of a ``high-velocity'' line.  However, the 
Gaussian components are only part of the high-velocity \hi\ phenomenon.  

The vast majority of the weak lines detected in the survey seem related both
spatially and kinematically to the lines known from previous work.
  Weak lines are not found outside the $V_{LSR}$ range of the strong lines, 
and there is no evidence that the weak lines 
merge into the Hubble flow.  
The integrated emission of the high-velocity components 
is skewed to negative LSR velocities but 
has a mean near zero when motion of the LSR about the Galactic Center
is removed.  

Most of the lines detected in this survey, at all values of  $N_{HI}$,  
are associated with complexes like  the Magellanic
Stream and Complex C.  The fraction of association may be 
as high as $90\%$.  
At the sensitivity of the survey the Magellanic Stream appears to 
extend at least $10\arcdeg$ to higher Galactic latitude 
than previously thought, and, near its tip, broadens in longitude 
as well.  Low-$N_{HI}$ lines associated with known 
complexes can be found far outside their traditional boundaries, 
 e.g., in the Magellanic Stream and Wright's cloud. 
The covering fraction of the sky increases as 
lower $N_{HI}$ lines are measured mainly because 
 the size of individual clouds grows.
 Our data thus suggest that,  
at the sensitivity level of the survey, 
most high-velocity clouds do not have sharp edges in \hi. 

Compact high-velocity clouds have received recent scrutiny because
they may be at a larger distance than the principle complexes 
\citep{bb99, bbc01}.  
There are a few low-$N_{HI}$ lines which appear spatially isolated
in the survey data (Table 9), but the majority of the emission appears 
to be extended, and in some directions  it is extended at 
low $N_{HI}$.  It would be quite valuable to observe  these with higher 
angular resolution to see if the material is clumped or 
remains fairly smooth.

Despite the abundance of low-$N_{HI}$ lines, 
their numbers are not sufficient to
contribute substantially to the mass of high-velocity \hi\ unless,
against the evidence, they are located far beyond the bright clouds.
This was the conclusion also of \citet{colgan}, and \citet{mls95}.
In a  set of compact, isolated, high-velocity clouds mapped 
at high angular resolution, the extended halos  contain most of 
the \hi\ flux \citep{bbc01}.  In this respect the compact clouds 
may have a different structure than the more general high-velocity 
emission.  

In contrast to spiral galaxies, whose \hi\ disks have sharp edges
and do not increase in size as observational sensitivity is 
increased \citep{vangorkom93}, the area covered by high-velocity 
\hi\ clouds is a strong function of the sensitivity of the observations 
(cf the current results with those of 
\citet{giovanelli}, \citet{wakk91}, and \citet{colgan}).
And yet, the high-velocity emission seems so concentrated into complexes 
that observation 
in only a few directions may easily 
 produce misleading impressions about its global 
properties (e.g., \citet{bowen95}).  For this reason we believe that many
more sight-lines must be studied in the UV before it can be determined 
if there is a statistical excess in the occurrence of, 
e.g., Mg II high-velocity lines
compared to 21cm high-velocity lines.

If the bright \hi\ lines which constitute the ``classic'' high-velocity
cloud population are merely the peaks in a more extended medium,
then the validity of statistical analyses based on cloud counts 
must be called into question.  A clumpy cloud may be counted several times
if observations are not sensitive enough to reveal the gas 
between the clumps.  Until there is a better way of specifying the
boundaries of a cloud than a noise level set by instrumentation, we believe
that statistical analyses of cloud counts are in danger of 
giving  misleading results. 

There are  high-velocity clouds which are 
 predominantly ionized  at $\sim 10^4$ K, there are clouds with 
 a significant component of mass at $T \sim 3 \times 10^5$ K, 
and there may even be clouds which are sources of soft X-ray emission 
\citep{sembach95, weiner96,  tufte98, kerp99, sembach99,sembach00}.  This 
suggests that the observed  low-$N_{HI}$ lines in some directions 
 might arise in  a trace neutral component of  an otherwise ionized cloud.  
The gas toward III Zw 2, where there 
is high-velocity Mg II absorption ($\S6.1$) but no detectable 
 21cm  \hi\ line,  may be in a similar state.
These considerations reenforce the view  that bright \hi\ clumps 
are merely one aspect of the high-velocity cloud phenomenon.

\acknowledgments

We thank the staff of the 140 Foot Telescope
 for their efforts throughout this project, and in particular 
the supervisor, George Liptak.  The 140 Foot Telescope
was retired in July 1999 after 26 years of service.
  We also thank Blair Savage, Brent Tully, Phil Maloney 
and John Hibbard for information, comments, and assistance.  
The referee made several stimulating comments which 
improved the paper significantly. 
The research of S.P-P. and V.J.U. at NRAO was supported by the NSF 
under an REU program grant.

\vfill \eject 


\clearpage
\begin{figure}
\epsscale{0.85}
\plotone{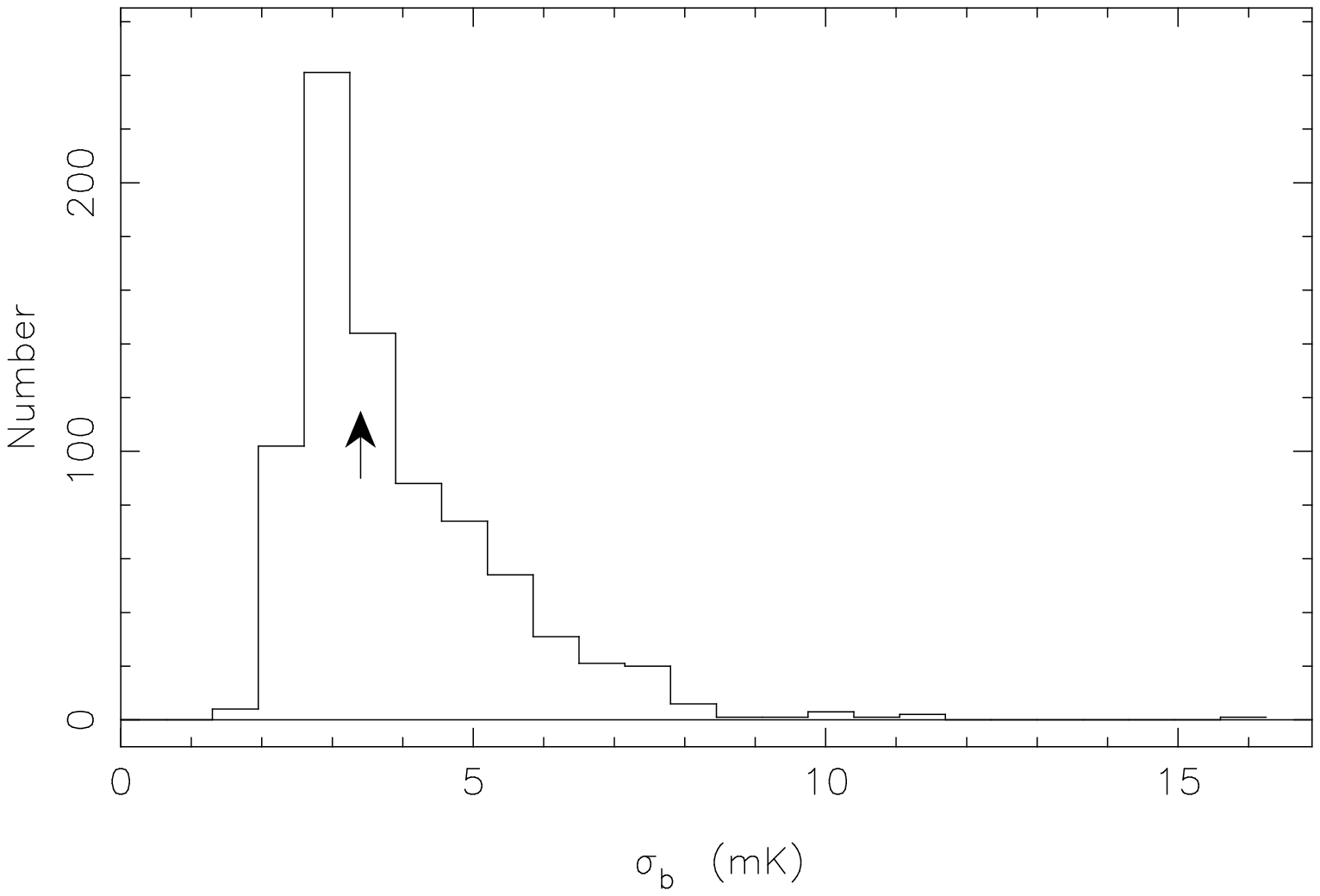}
\caption{The distribution of the rms 
baseline noise in the final position-switched
spectra.  The median value, 3.4 mK, is shown with an arrow.}
\end{figure}
\clearpage

\clearpage
\begin{figure}
\epsscale{1.05}
\plotone{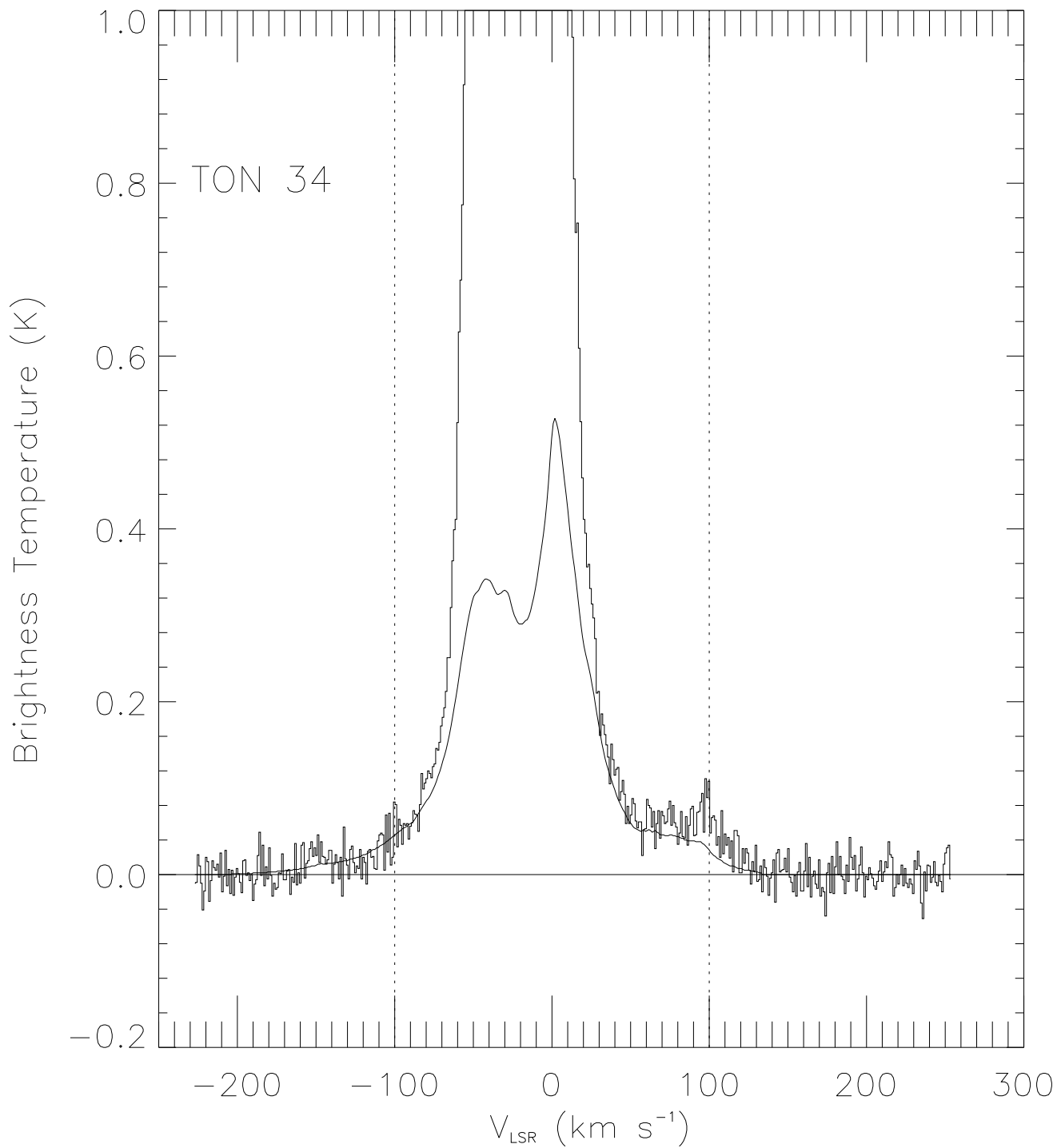}
\caption{A portion of a frequency-switched \hi\ spectrum 
toward Ton 34  at $\ell,b = 203\fdg0,+56\fdg5$ illustrating stray radiation.
Superimposed on the observed 140 Foot spectrum is a smooth curve showing 
the amount of the detected signal attributed to stray radiation
 which has entered the telescope's
sidelobes.  The long wing to positive velocity is almost entirely 
spurious, though
more sensitive position-switched data show that the small 
excess emission above the stray near V$=+100$ km~s$^{-1}$ is actually
real.  The observations have been clipped at 1.0 K for display; 
the corrected peak T$_b$ of the line is 4.5 K.  
}
\end{figure}
\clearpage

\clearpage
\begin{figure}
\epsscale{0.8}
\plotone{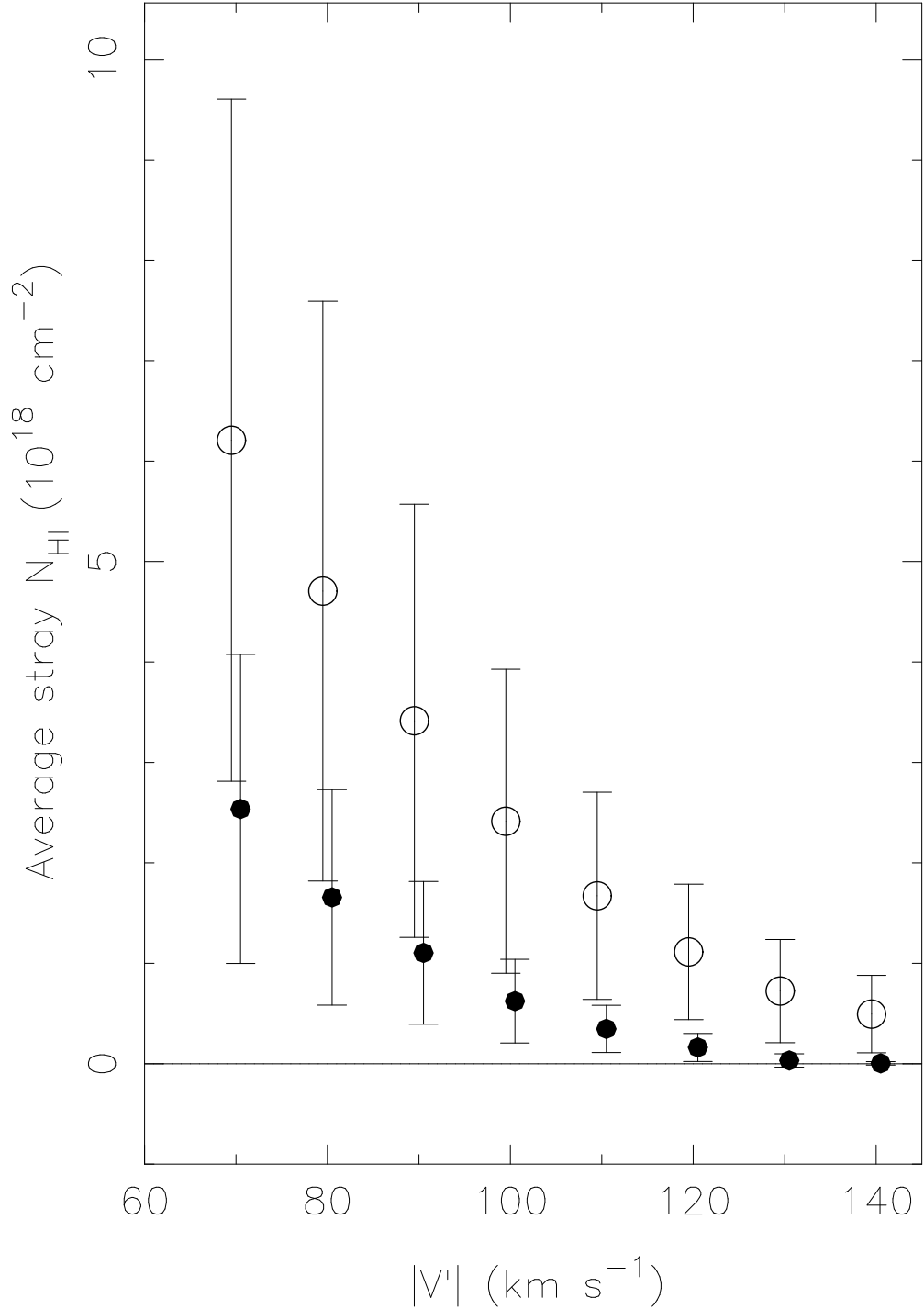}
\caption{The average amount of stray radiation, expressed as an 
equivalent $N_{HI}$, that was removed from the $\sim 100$ frequency-switched
spectra in the survey. 
For each value of $V'$ the Figure shows the  amount of stray 
radiation at  greater (or lesser) LSR 
velocities.  
 Stray radiation at negative velocities is 
indicated with an open circle; at positive velocities with a filled circle.
This Figure shows that  the average frequency-switched 
spectrum in the survey had a spurious wing at 
$V_{LSR} < -100 $ km~s$^{-1}$
with $N_{HI} \sim 2 \times 10^{18}$, and 
a spurious wing at  $V_{LSR} > +100 $ km~s$^{-1}$ 
with  $N_{HI} \sim 0.6 \times 10^{18}$ cm$^{-2}$. 
}
\end{figure}
\clearpage

\clearpage
\begin{figure}
\epsscale{1.02}
\plotone{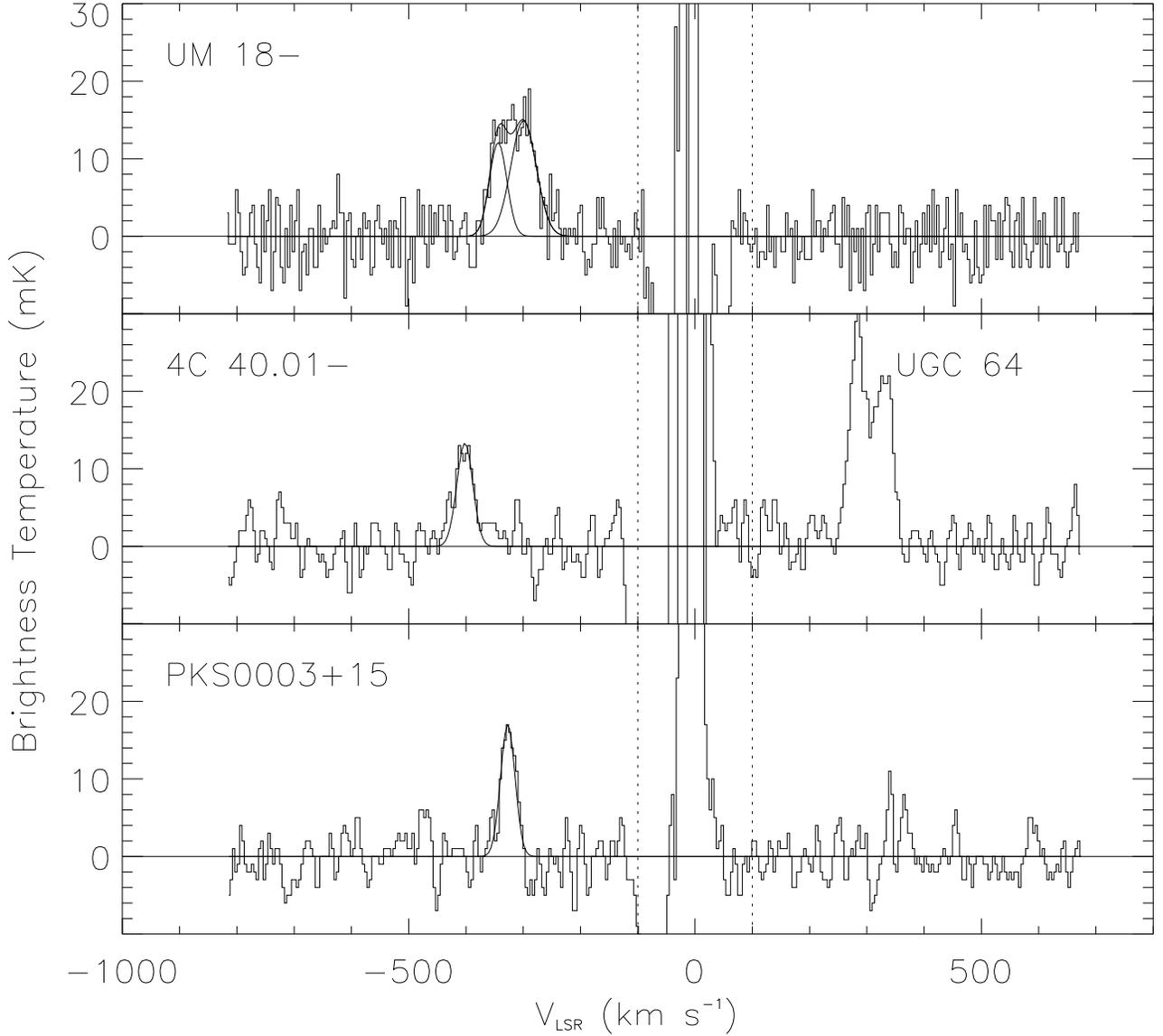}
\caption{Sample position-switched spectra showing the data quality and
examples of the weaker \hi\  emission lines detected in the survey.  
Vertical lines
at $\pm100$ km~s$^{-1}$ mark the adopted limits for high-velocity
gas.  Emission at $|V_{LSR}| < 100$ km~s$^{-1}$ is often nearly canceled
in the position-switching.  The lower two spectra have been Hanning
smoothed.  The top panel shows two lines 
separated by 44 km~s$^{-1}$ near $-300$ km~s$^{-1}$ 
with $N_{HI} = 8.3 \times 10^{17}$
and $1.5 \times 10^{18}$ cm$^{-2}$.  The central panel has a line
from a high-velocity cloud at $-400$ km~s$^{-1}$ with 
$N_{HI} = 8.5 \times 10^{17}$ cm$^{-2}$,
 and, near $+300$ km~s$^{-1}$,  \hi\ emission from the galaxy UGC 64 
which lies serendipitously in the antenna beam.  In the lower panel, 
the line at $V_{LSR} = -327$ km~s$^{-1}$ has a FWHM of 30 km~s$^{-1}$,
equal to the median of all detected lines, and 
$N_{HI} = 9.8 \times 10^{17}$ cm$^{-2}$.  The rms noise in the spectra is 
 3.5 mK, 2.7 mK, and 2.4 mK, upper to lower. The median completeness level 
of the survey is $8 \times 10^{17}$ cm$^{-2}$, similar to the $N_{HI}$
in most  of the high-velocity lines shown here.
}
\end{figure}
\clearpage

\clearpage
\begin{figure}
\epsscale{1.05}
\plotone{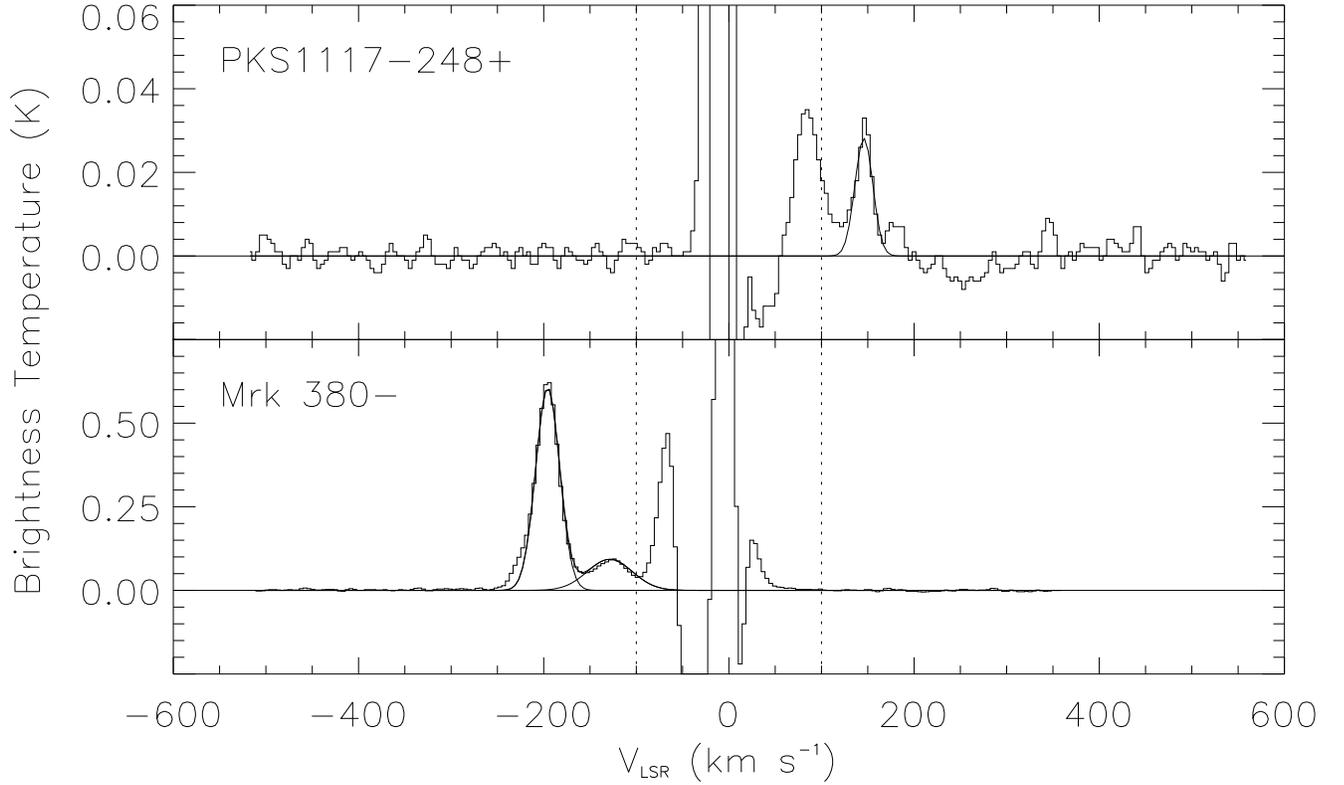}
\caption{
The central portions of two smoothed spectra showing wings and blended lines.  
The temperature scales differ by more than an order of
magnitude.  The upper panel has a Gaussian component at +145 km~s$^{-1}$ 
with $N_{HI} = 1.25 \times 10^{18}$ cm$^{-2}$ and additional 
emission at a lower velocity which is treated as a ``wing'' 
because its shape at $V_{LSR} \lesssim 80 $ km~s$^{-1}$ arises from 
cancellation in the position-switching and hence  Gaussian parameters 
would have no physical significance.  In the lower panel 
the \hi\ at V$ < -100$ km~s$^{-1}$  is treated as two 
individual Gaussians plus some additional emission.  
  The rms noise in the upper spectrum is 2.2 mK; 
in the lower spectrum it is 3.5 mK. An instrumental problem caused the
apparent absorption in the upper panel at $+250$ km s$^{-1}$ (see 
$\S2.4$). 
}
\end{figure}
\clearpage

\clearpage
\begin{figure}
\epsscale{1.05}
\plotone{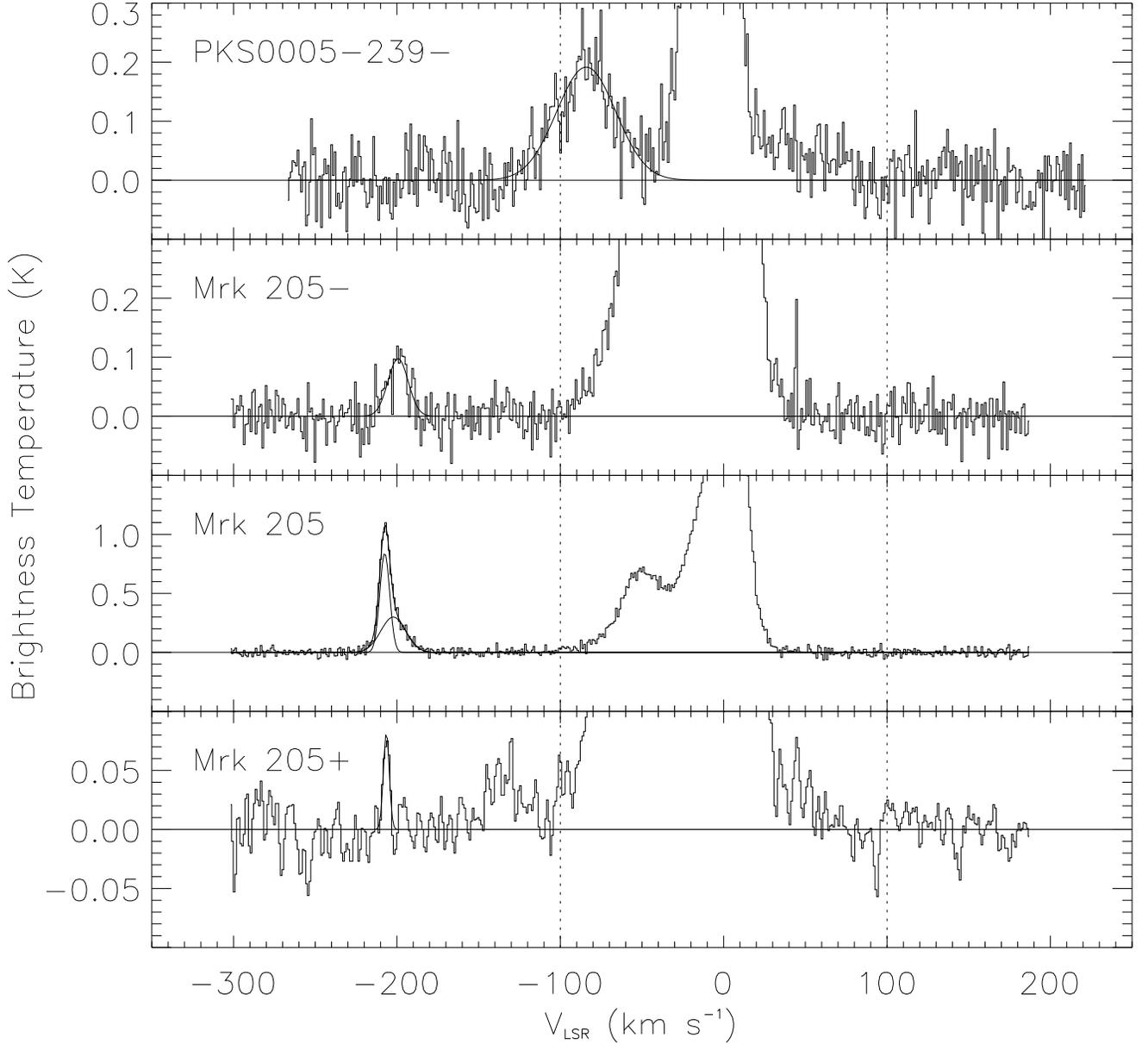}
\caption{Representative frequency-switched spectra.  
Only the bottom one has been smoothed.
All have been corrected for stray radiation.  There is  
weak interference in the Mrk $205-$ spectrum near $+40$ km~s$^{-1}$.
The apparent emission at $-140$ km~s$^{-1}$ toward Mrk 205+ could not be
confirmed in the position-switched data and may be residual stray
radiation which the deconvolution process did not remove.
The line toward Mrk 205+ is the narrowest in the survey, with 
$\Delta v = 4.2 \pm 0.8$ km s$^{-1}$, and $N_{HI} = 6.7 \times 
10^{17}$. 
}
\end{figure}
\clearpage

\clearpage
\begin{figure}
\epsscale{0.7}
\plotone{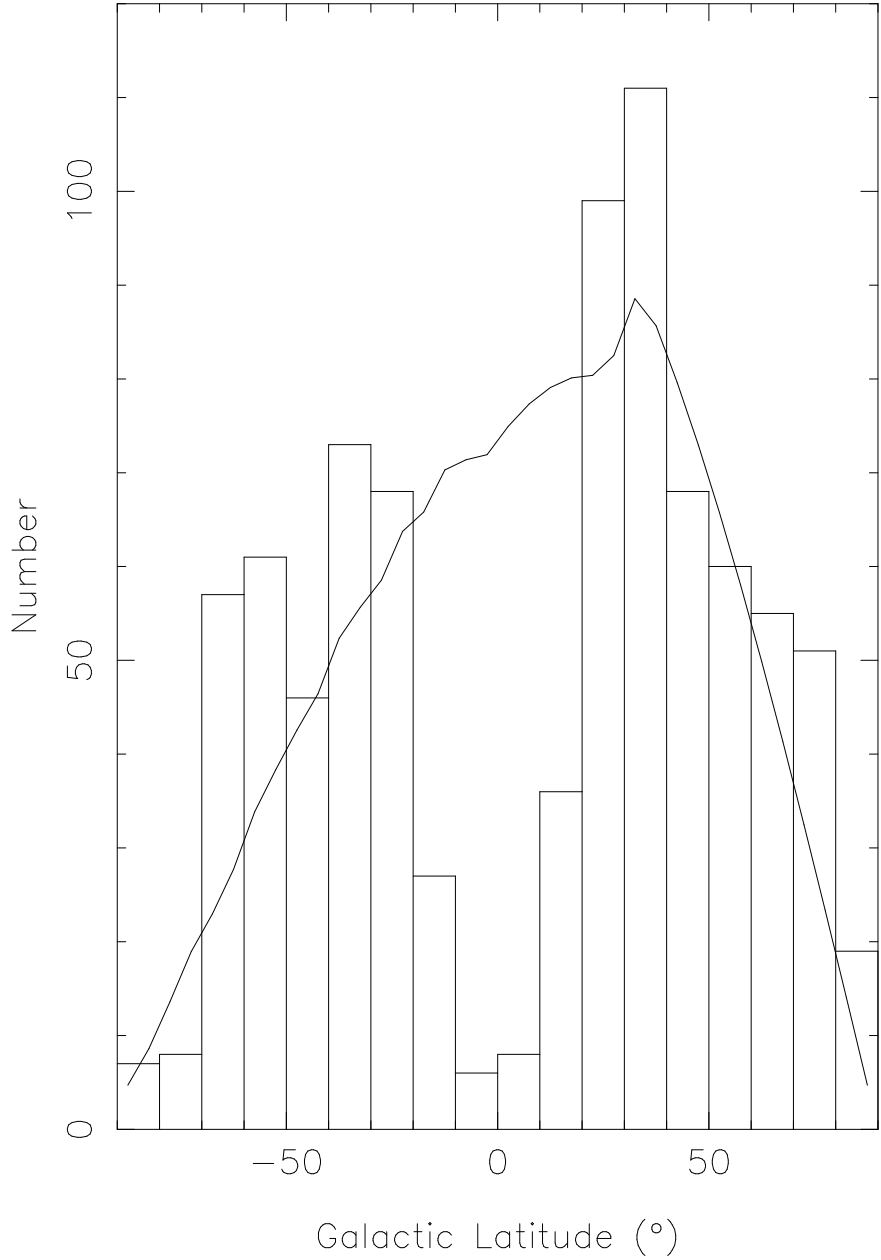}
\caption{The distribution of observed Galactic latitudes, with
a curve proportional to the fraction of the sky 
at each latitude which lies
at $\delta \geq -30\arcdeg$.  The survey has reasonably uniform coverage
in latitude except at $|b|<20\arcdeg$, where the selection criterion
produces a bias against directions with high extinction.
}
\end{figure}
\clearpage

\clearpage
\begin{figure}
\epsscale{0.7}
\plotone{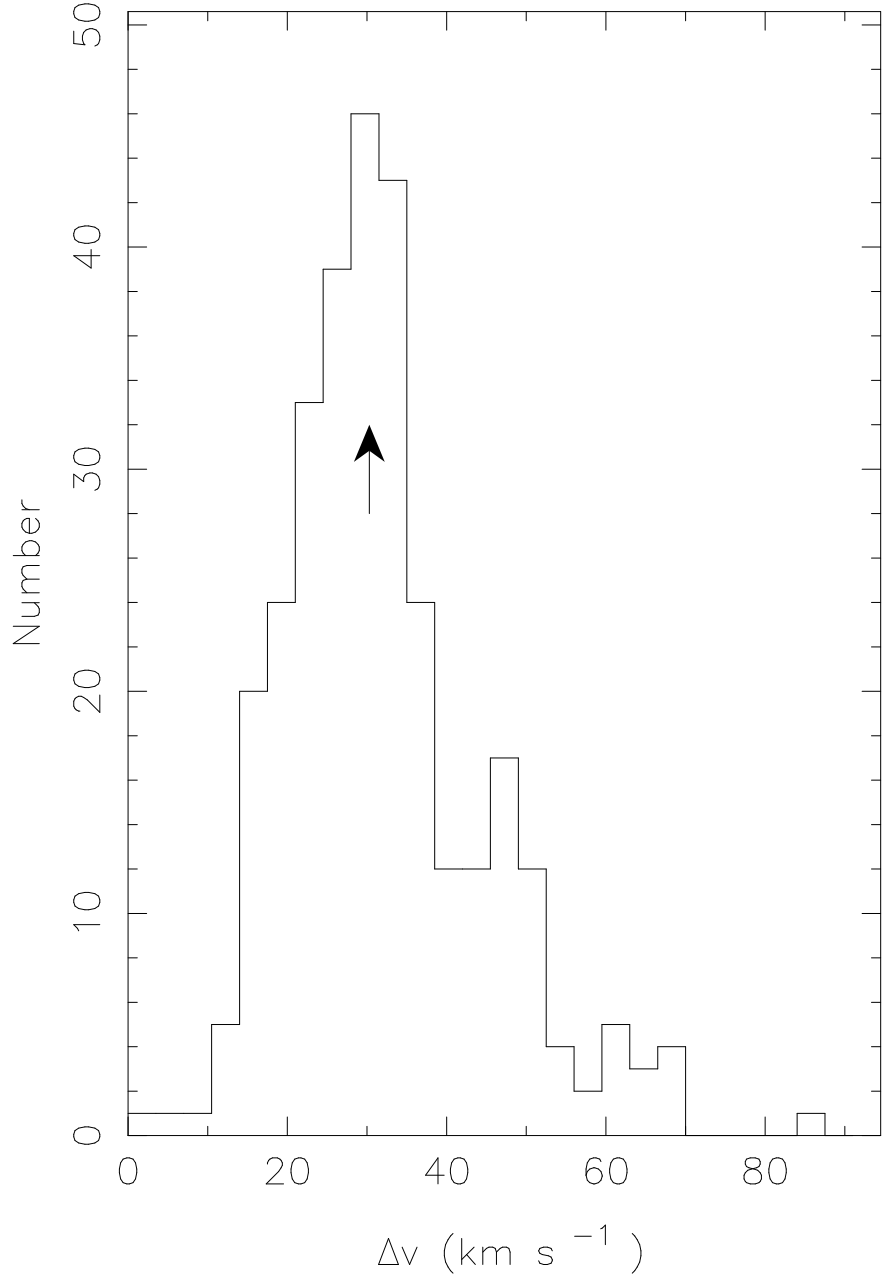}
\caption{The distribution of linewidths after correction in quadrature for
the width of the spectrometer channels.  The median value of 30.3 
km~s$^{-1}$ is marked with an arrow.
}
\end{figure}
\clearpage

\clearpage
\begin{figure}
\epsscale{0.95}
\plotone{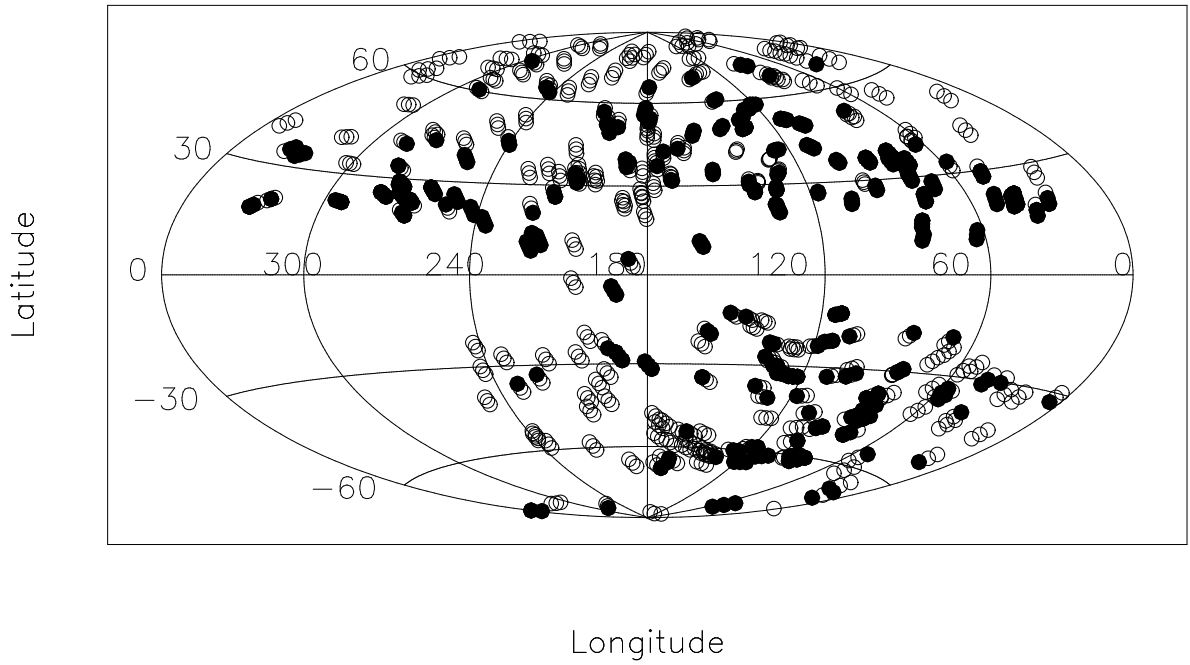}
\caption{The survey directions in an Aitoff projection centered
on $\ell,b=180\arcdeg,0\arcdeg$.  Open symbols mark directions without 
high-velocity \hi,  filled symbols mark detections of \hi\ 
at $|V_{LSR}|>100$ km~s$^{-1}$.
}
\end{figure}
\clearpage

\clearpage
\begin{figure}
\epsscale{0.95}
\plotone{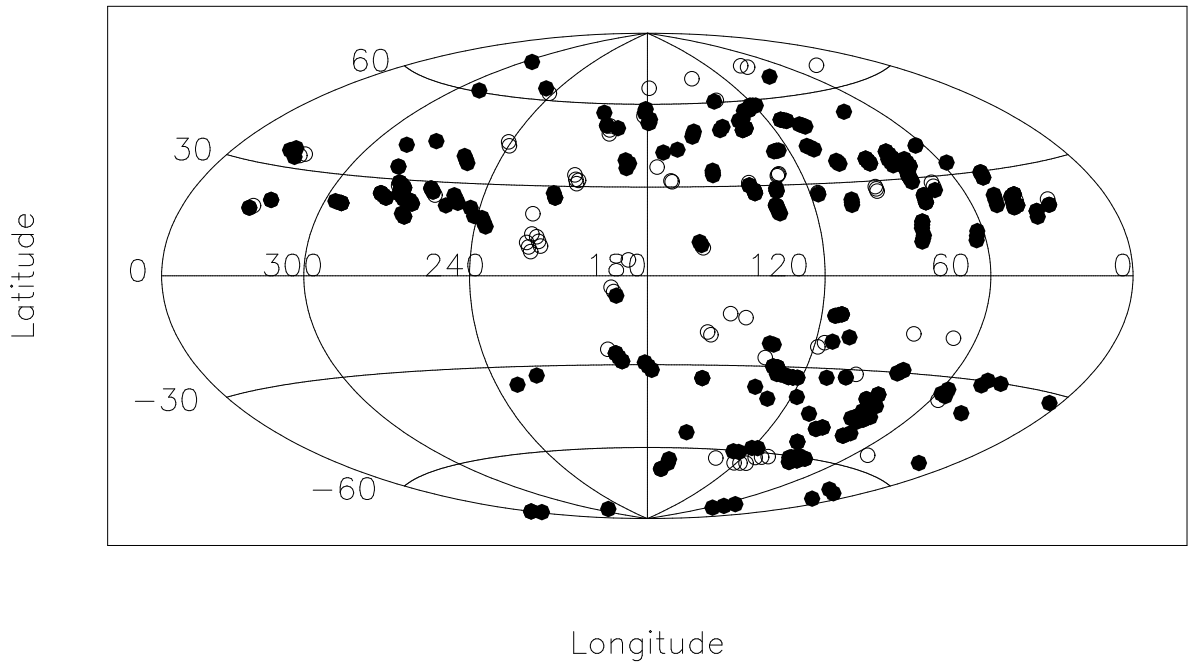}
\caption{An Aitoff projection centered
on $\ell,b=180\arcdeg,0\arcdeg$ showing directions where high-velocity 
\hi\ was detected.  Open symbols identify lines that are wings
blended with low-velocity emission, and solid symbols identify 
\hi\ lines  well-separated from low-velocity
emission which can be characterized by a Gaussian function.
}
\end{figure}
\clearpage

\clearpage
\begin{figure}
\epsscale{0.85}
\plotone{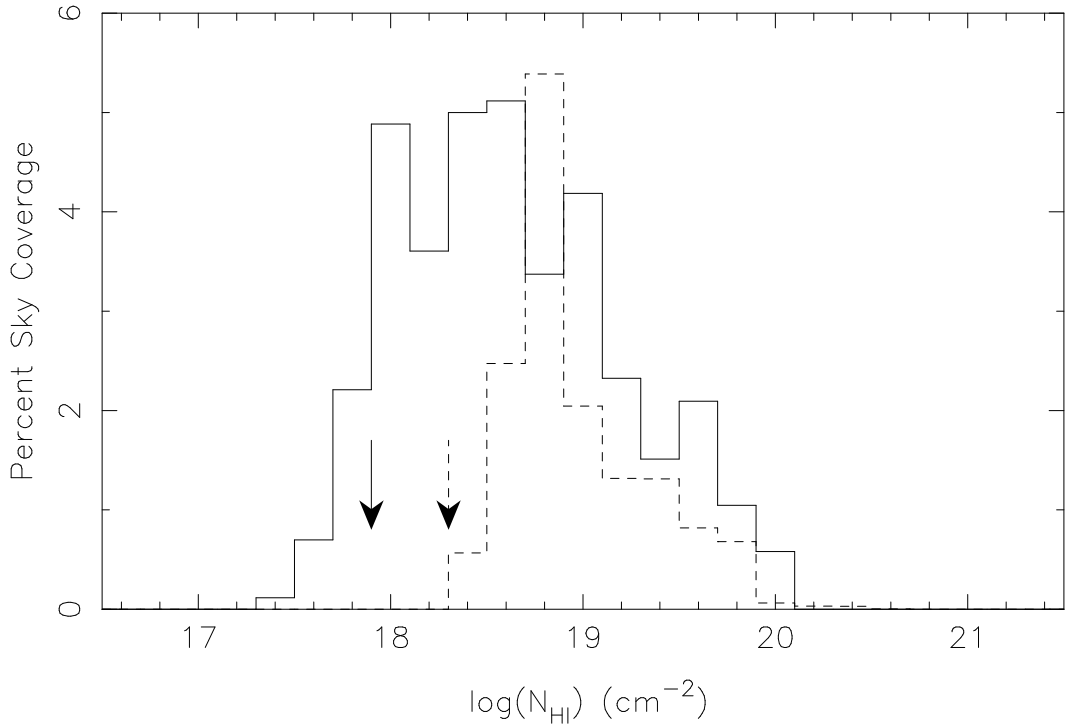}
\caption{The solid line is the percentage of the sky covered by 
high-velocity \hi\ emission in each interval of log($N_{HI}$) from 
the data of this survey.  The solid arrow marks the estimated 
median completeness
limit of the survey: $8 \times 10^{17}$ cm$^{-2}$.  The dashed line 
is the percentage sky coverage derived from the compilation of 
\citet{wakk91} and the dashed arrow is the estimated completeness
level of that compilation.  Summed over all 
values of $N_{HI}$, 37\% of the sky is  covered with 
detectable high-velocity \hi.
}
\end{figure}
\clearpage

\clearpage
\begin{figure}
\epsscale{0.95}
\plotone{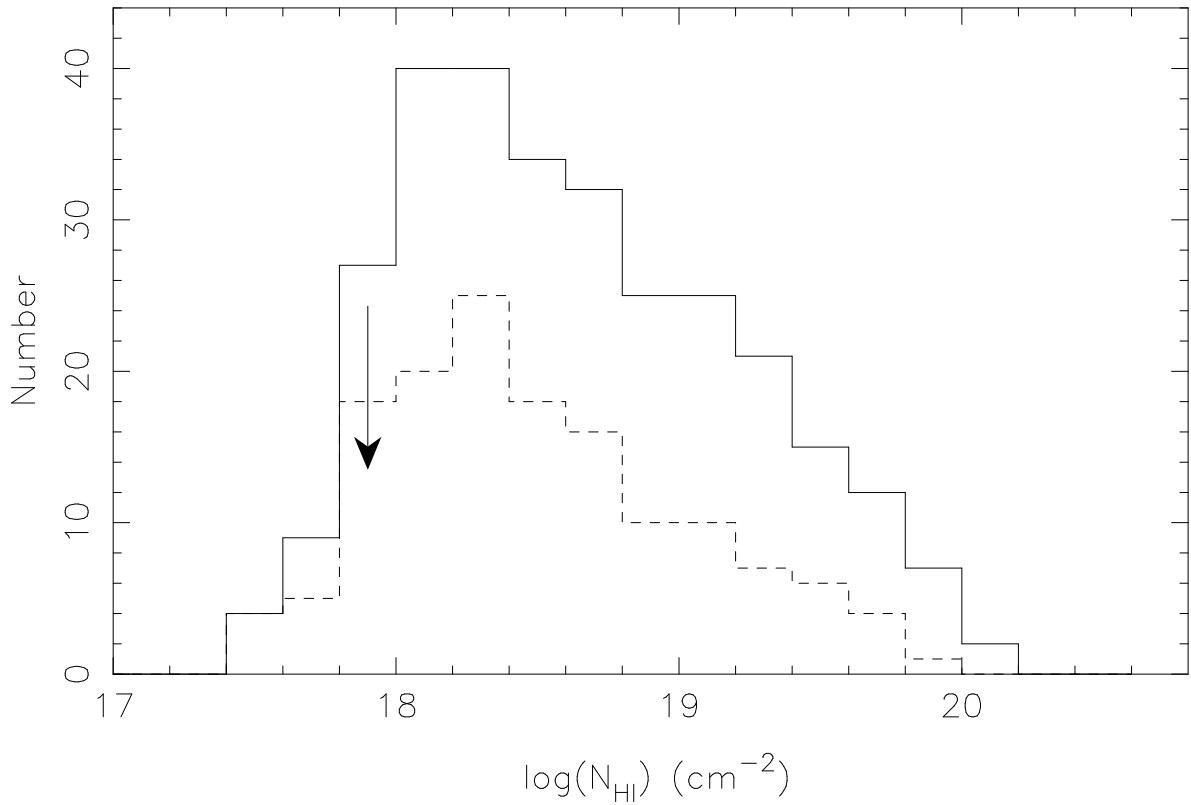}
\caption{The number of  Gaussian high-velocity
\hi\ lines detected in the survey as a function of their log($N_{HI}$).
The solid curve is for all lines with $|V_{pk}| \geq 
100$  km~s$^{-1}$, and the dashed curve is for the higher velocity lines 
with $|V_{pk}| \geq 150$ km~s$^{-1}$ only.  The arrow shows the completeness
level of the survey. 
}
\end{figure}
\clearpage

\clearpage
\begin{figure}
\epsscale{0.85}
\plotone{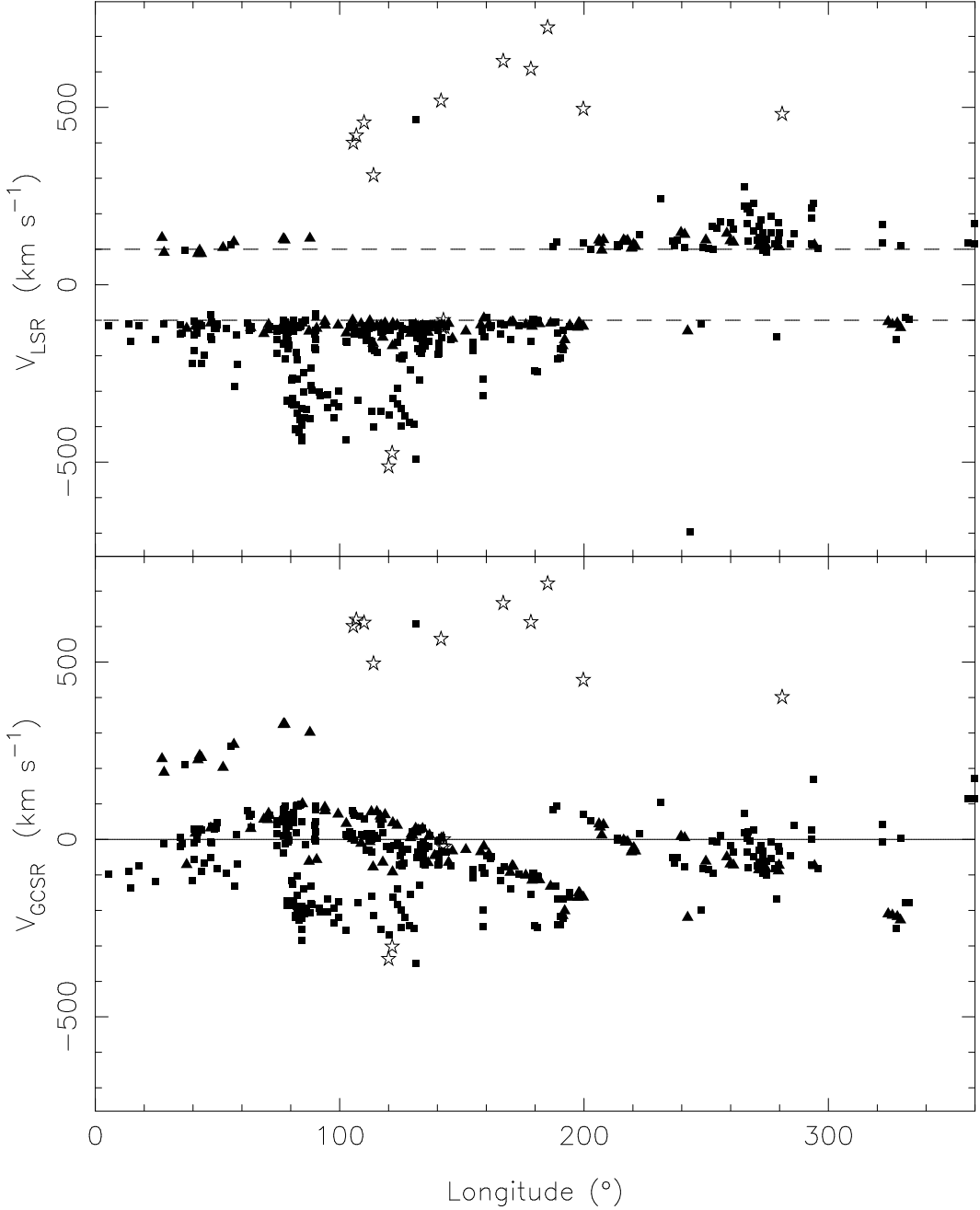}
\caption{Velocity of high-velocity \hi\ lines 
with respect to the LSR, and to the Galactic Center, 
plotted against Galactic 
longitude.  Filled squares are Gaussian components, 
filled triangles are line
wings, and open stars are known galaxies which were detected 
by accident during the survey.
The dashed lines in the upper panel are at $\pm100$ km~s$^{-1}$ and show the 
 adopted $V_{LSR}$ limit for high-velocity gas.  There is a 
distinct transition 
between the high-velocity \hi\ clouds and the positive-velocity 
galaxies which are part of the Hubble flow 
(the filled squares at V$_{LSR}=+465$  and $-700$ km~s$^{-1}$ are from 
very weak lines which may not be real).  Much of the 
pattern of $V_{LSR}(\ell)$ is due to Galactic rotation.
}
\end{figure}
\clearpage

\clearpage
\begin{figure}
\epsscale{0.95}
\plotone{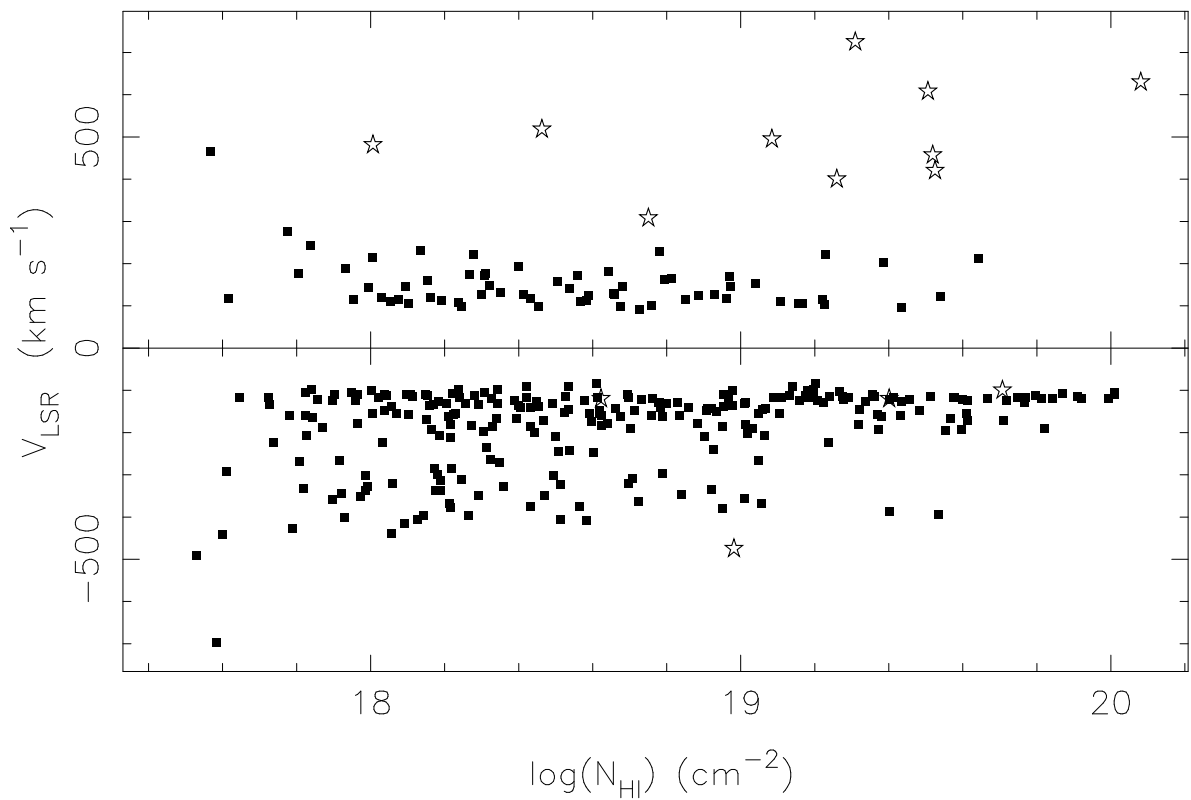}
\caption{The LSR velocity of the Gaussian component 
high-velocity \hi\ lines (Table 4)  vs.~their $N_{HI}$, 
where galaxies detected in the survey are shown as open stars. 
The weakest lines have kinematics similar to all but the very 
strongest lines.  Values of $N_{HI}$ for the galaxies are lower limits 
because the observations were not necessarily centered on those objects.
There is a distinct separation between the high-velocity \hi\ lines 
and the lines from galaxies which, 
apart from  M31 at $V_{LSR} \approx -500$ km s$^{-1}$, 
  are part of the Hubble flow.
}
\end{figure}
\clearpage

\clearpage
\begin{figure}
\epsscale{0.7}
\plotone{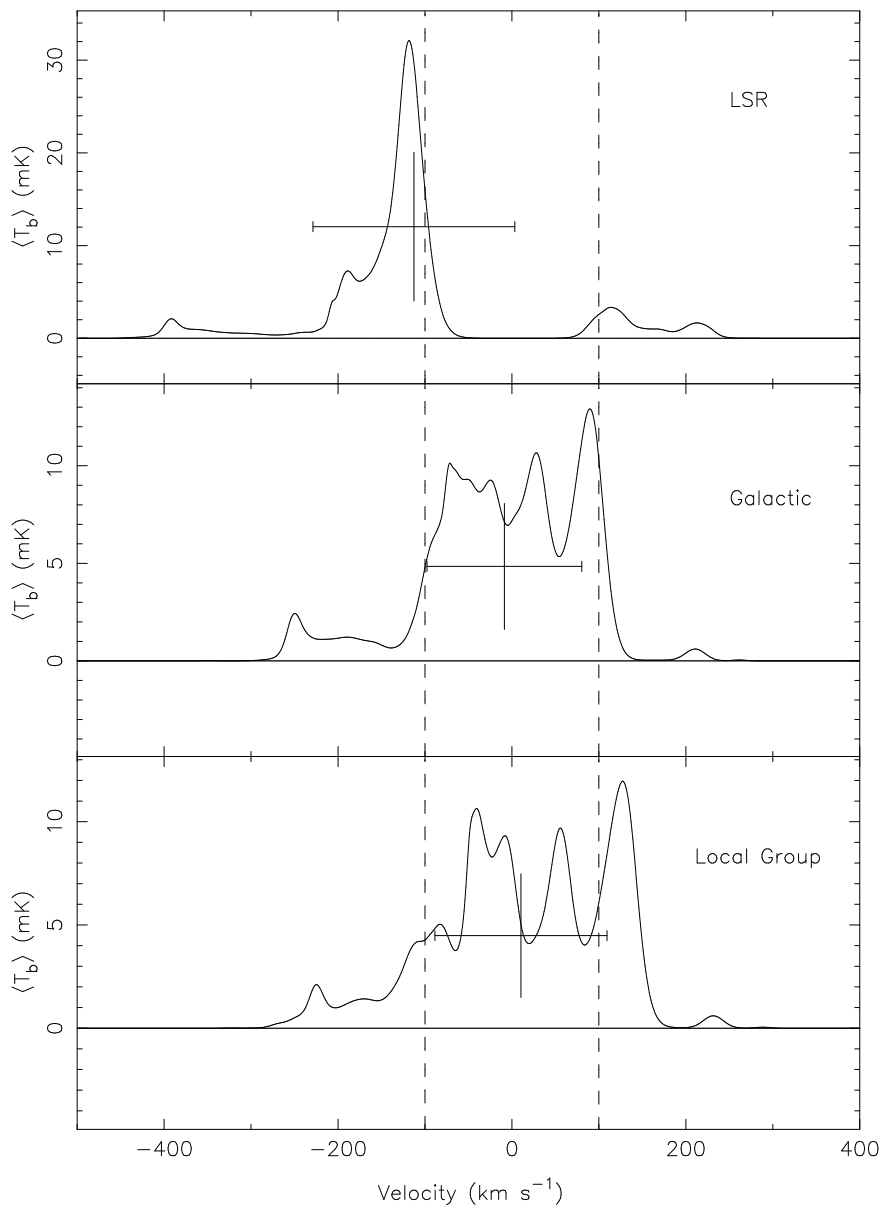}
\caption{The total 21cm \hi\ emission of the high-velocity Gaussian components
summed with respect to the LSR (top panel), 
with respect to the Galactic Center standard
of rest (center panel), and with respect to the
Local Group standard of rest (lower panel).
The dashed lines are at $\pm 100$ km~s$^{-1}$.  The cross in each spectrum
is centered at the mean velocity 
 of the emission and the horizontal bar is $\pm 1$ standard 
deviation about the mean.  The most negative velocity emission 
comes from the Magellanic Stream.  The emission summed in the 
Galactic Center system has a mean velocity closest to zero and 
the smallest dispersion of the three (see Table 6).  
}
\end{figure}
\clearpage

\clearpage
\begin{figure}
\epsscale{0.85}
\plotone{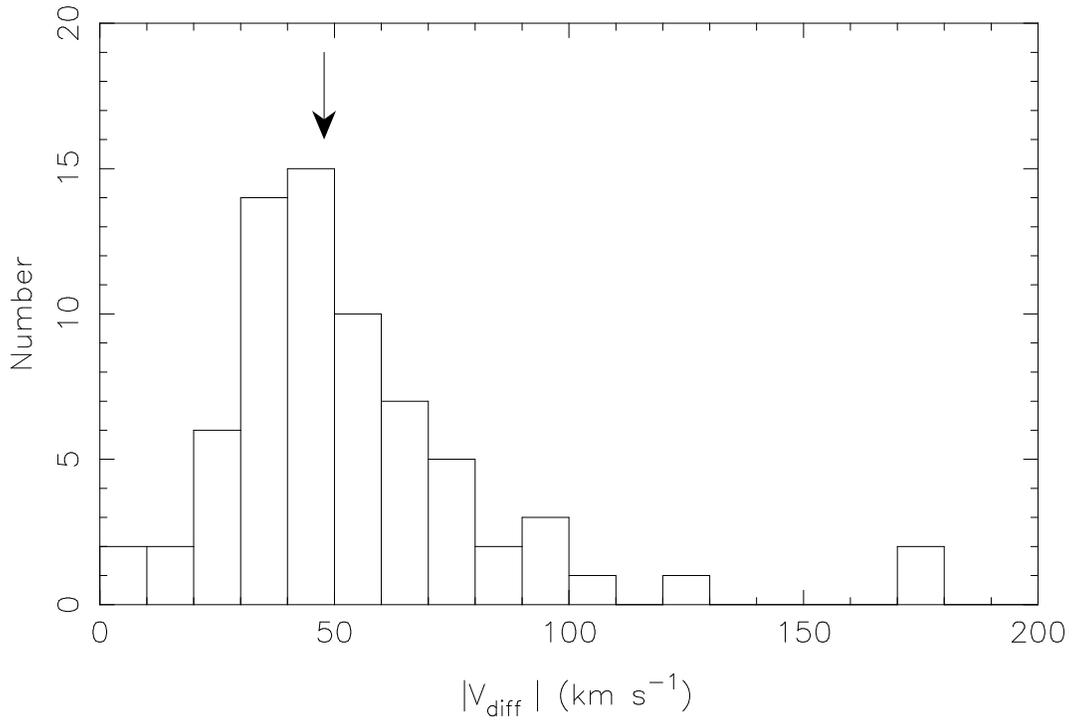}
\caption{The distribution of the difference in $V_{LSR}$ 
between high-velocity lines which appear in the same spectrum.
The arrow marks the median, 48 km~s$^{-1}$. There are few 
lines with very  small values of  $|V_{diff}|$,
 which might be seen  from clouds having a core-halo structure.
}
\end{figure}
\clearpage

\clearpage
\begin{figure}
\epsscale{0.95}
\plotone{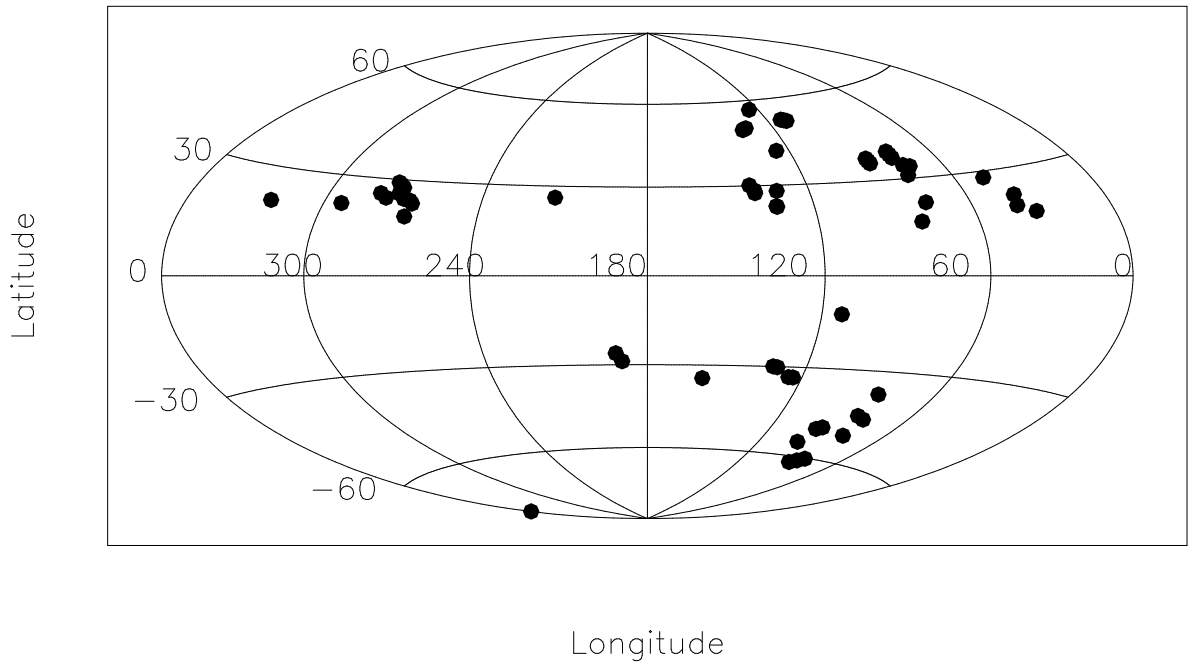}
\caption{The location of directions with multiple high-velocity \hi\ 
lines in a single spectrum.  Most of the multiple lines 
appear to arise from internal structure in large high-velocity complexes.
The Magellanic Stream, Complex C, 
 Complex A, and Wright's Cloud are all represented here.}
\end{figure}
\clearpage

\clearpage
\begin{figure}
\epsscale{0.95}
\plotone{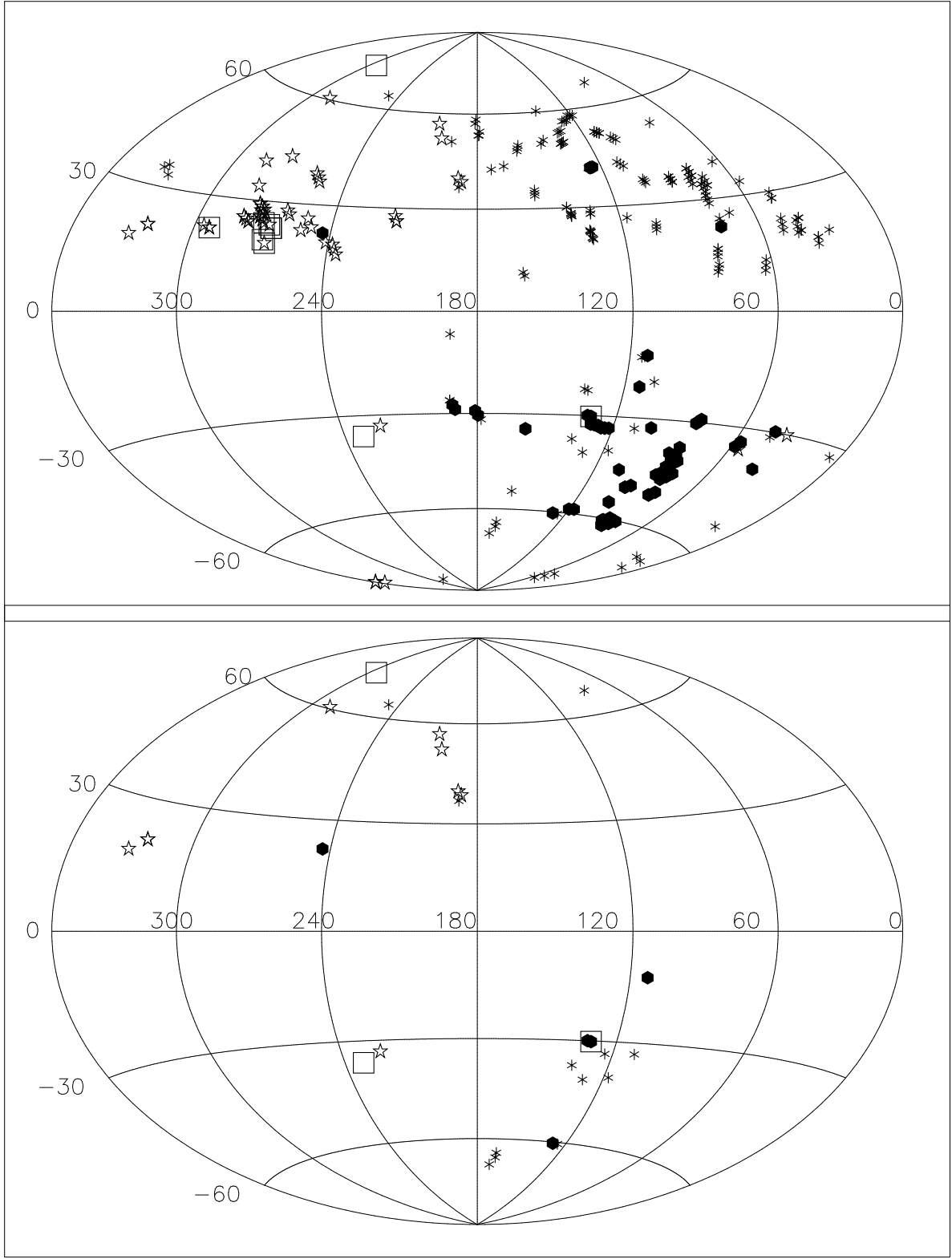}
\caption{Upper panel:  All  high-velocity Gaussian components.  Filled
symbols mark lines with $V_{LSR}~<~-200$ km~s$^{-1}$, asterisks mark 
V$_{LSR}$ in the range
$-200$ to  $ -100$ km~s$^{-1}$, open stars $ +100$ to $+200$ 
km~s$^{-1}$, and open rectangles $>+200$ km~s$^{-1}$.  The association 
of lines into coherent complexes is evident. 
Lower panel: Same as the upper panel only for lines which could not be 
assigned to one of the large complexes. 
}
\end{figure}
\clearpage

\clearpage
\begin{figure}
\epsscale{1.0}
\plotone{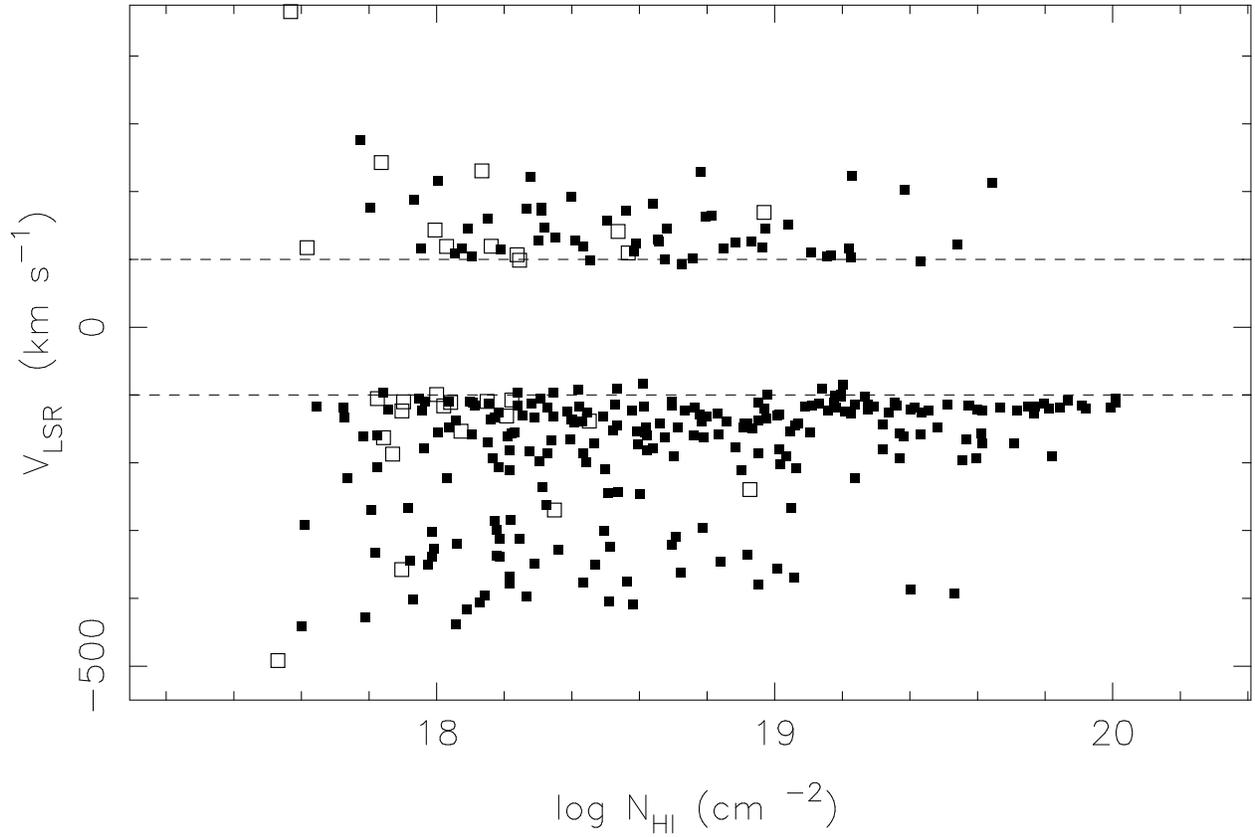} 
\caption{The LSR velocity vs. $N_{HI}$ of the 
Gaussian components from Table 4, where
filled symbols identify lines associated or possibly associated 
with a large high-velocity complex, 
and open symbols are lines not associated with a known complex.
The reality of the lines at $+465$ and $-500$ km~s$^{-1}$ is suspect,
and there is one suspect line with  $V_{LSR} \sim -700$ km~s$^{-1}$ 
and log($N_{HI}) = 17.6$  which is not plotted (see $\S3.4$). 
}
\end{figure}
\clearpage

\clearpage
\begin{figure}
\epsscale{0.90}
\plotone{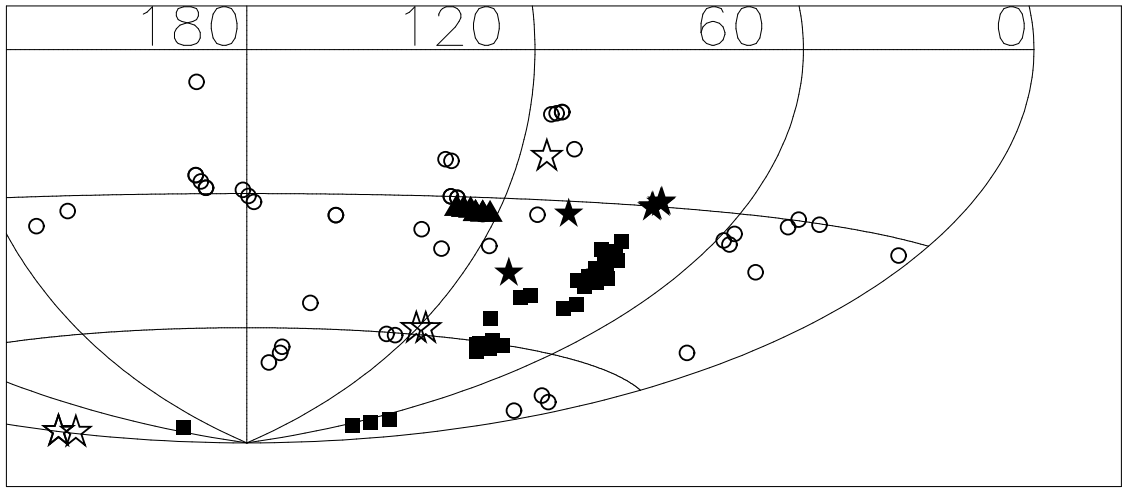}
\caption{An Aitoff projection in Galactic coordinates 
of high-velocity \hi\ lines near the 
Magellanic Stream.  
  Lines within the traditional borders of the Stream
are filled squares, and those probably unrelated to the Stream 
 because of their velocity are open
circles.  The lines marked by filled stars show emission detected
in this survey which traces an 
 extension of the Stream to higher latitude and longitude.
These weak lines, listed as MS* in Table 4, 
 would not have been detected in  previous surveys.  The open
stars are lines which might be part of  the Stream, but their association
is less certain (MS?).  Filled triangles mark lines from 
Wright's cloud near M33, which is considerably more extended in our
data than in earlier studies.
}
\end{figure}
\clearpage

\clearpage
\begin{figure}
\epsscale{0.90}
\plotone{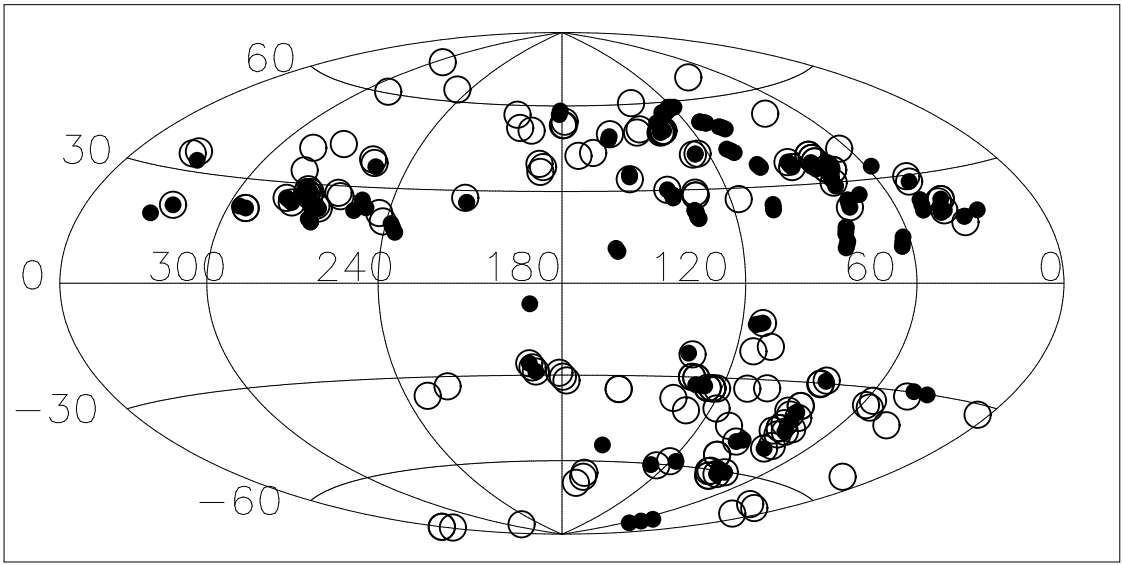}
\caption{ All the high velocity lines from Table 4 sorted by their 
total $N_{HI}$.  The brightest half of the lines ($>3.5 \times 10^{18}$ 
cm$^{-2}$) are the filled symbols and the faintest half the open 
symbols. Although there are some chance superpositions of clouds 
at different velocities, lines in similar parts of the sky 
tend to have similar velocities (Fig.~18), and  in many instances the faint 
lines surround and connect  bright lines.  
}
\end{figure}
\clearpage

\clearpage
\begin{figure}
\epsscale{0.70}
\plotone{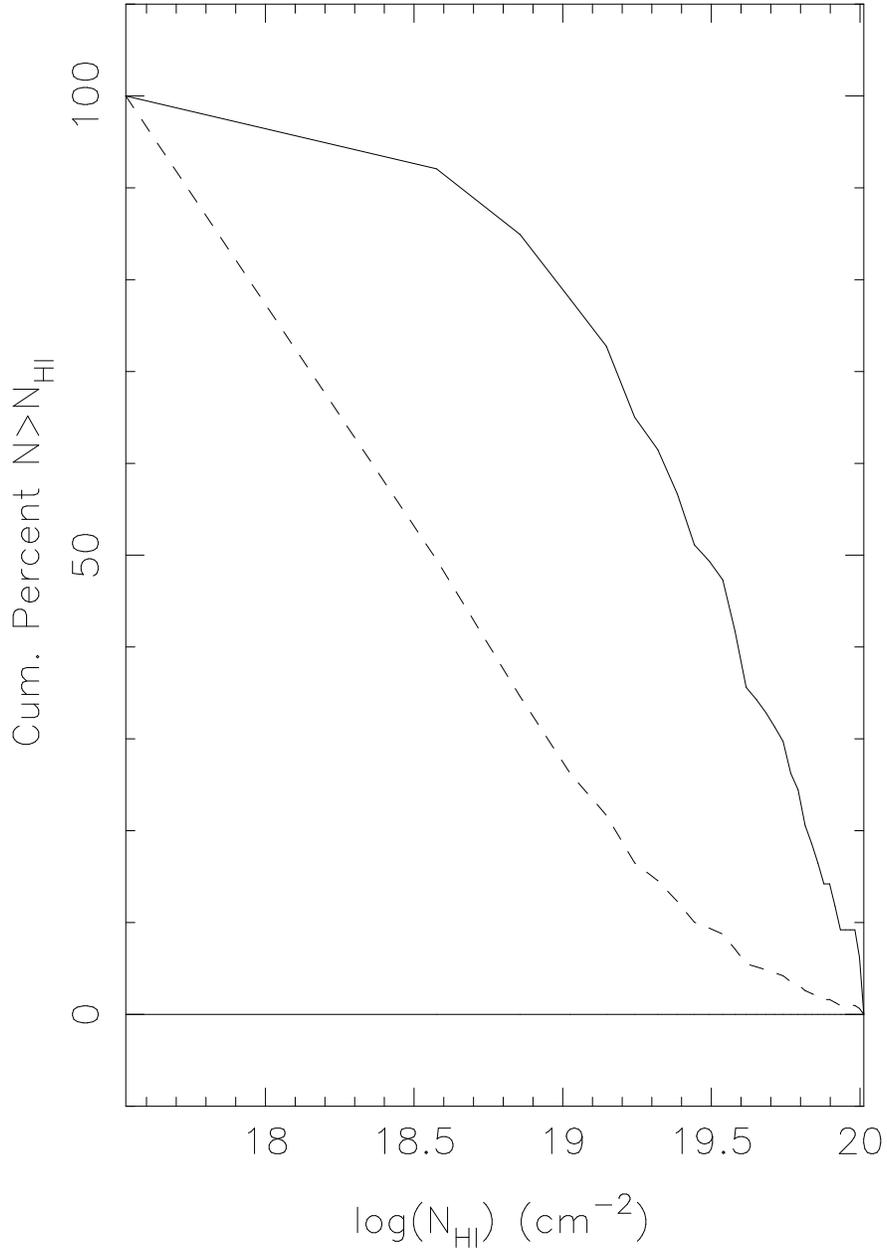}
\caption{ The cumulative distribution of the numbers of detected 
high-velocity lines (dashed curve) and  the total  $N_{HI}$ 
that they contain (solid curve) for all lines above the given 
value of  log($N_{HI}$).  Most of the lines have a  low 
column density, but most of the atoms reside in lines with the 
highest  $N_{HI}$.
}
\end{figure}
\clearpage


\begin{deluxetable}{lrrrrl} 
\tabletypesize{\footnotesize}
\tablecolumns{5} 
\tablenum{1}
\tablewidth{0pc} 
\tablecaption{Observed Central Positions}
\tablehead{ 
     \colhead{Object Name} &
     \colhead{$\alpha$(B1950)} &
     \colhead{$\delta$(B1950)} &
     \colhead{$l$ } &
     \colhead{$b$ } &\\

}
\tablecomments{Units of right ascension are hours, minutes, and
seconds, and units of declination are degrees, arcminutes, and
arcseconds.  Units of Galactic longitude and latitude are degrees.}

\startdata 

UM 18          &00 02 46&05 07 29 &102.310&$-$55.612 & \\
PKS 0003+15    &00 03 25&15 53 08 &107.318&$-$45.325 & \\
PKS 0005$-$239   &00 05 27&$-$23 55 59& 50.039&$-$79.599 & \\
III ZW 2       &00 07 55&10 42 02 &106.970&$-$50.620 & \\
4C 40.01       &00 10 54&40 34 57 &115.241&$-$21.444 & \\
Q 0022+0015    &00 22 10&00 15 44 &108.786&$-$61.597 & \\
NAB 0024+22    &00 24 38&22 25 24 &115.663&$-$39.830 & \\
5C 03.44       &00 34 54&39 21 42 &120.037&$-$23.157 & \\
\enddata 

\tablenotetext{a}{Flanking positions offset $13\fm5$ in right ascension.}
\tablenotetext{b}{Flanking position at lower right ascension only.}

\end{deluxetable} 


%

\begin{deluxetable}{cccccc}
\tablecolumns{6}
\tablenum{2}
\tablewidth{0pt}
\tablecaption{Summary of Experimental Parameters.}
\tablehead{
\colhead{Mode} & \colhead{Hann\tablenotemark{a}} &
\colhead{ch.~width\tablenotemark{b} }  
& \colhead{median $\sigma_b$} & 
\colhead{Vel. coverage} & 
\colhead{Numb.}  \\
\colhead{} & \colhead{} &
\colhead{(km s$^{-1}$)}  & \colhead{(mK)} & \colhead{(km s$^{-1}$)}  \\
} 
\startdata
Pos-Sw & Y &  10.0 & 3.1 & 1800 & 574\\
Pos-Sw & N & 5.0 & 5.1 & 1800 & 221 \\
Freq-Sw & Y & 2.5 & 10.6 & 450 & 16 \\
Freq-Sw & N & 1.25 & 22.5 & 450 & 49 \\
\enddata
\tablenotetext{a}{Is spectrum  Hanning smoothed?} 
\tablenotetext{b}{Effective velocity width of a single spectral channel.}
\end{deluxetable}

\vfill \eject


\begin{deluxetable}{rlrrrrrrrrrrll} 
\tabletypesize{\scriptsize}
\rotate
\tablecolumns{14} 
\tablenum{3}
\tablewidth{0pc} 
\tablecaption{Detections of High-Velocity Hydrogen }
\tablehead{ 
  \colhead{\#} & 
  \colhead{Object} &
  \colhead{$\alpha$(B1950)} &
  \colhead{$\delta$(B1950)} &
  \colhead{$l$} &
  \colhead{$b$} &
  \colhead{T$_{\rm peak}$} &
  \colhead{$\Delta$v} &
  \colhead{V$_{\rm LSR}$} &
  \colhead{$\sigma_{\rm g}$} &
  \colhead{$\sigma_{\rm b}$} &
  \colhead{N$_{HI}$} &
  \colhead{$<$V$>$} &
  \colhead{Notes}\\

  \colhead{} & 
  \colhead{} &
  \colhead{} &
  \colhead{ } &
  \colhead{(deg)} &
  \colhead{(deg)} &
  \colhead{(mK)} &
  \colhead{($\rm km\ s^{-1}$)} &
  \colhead{($\rm km\ s^{-1}$)} &
  \colhead{(mK)} &
  \colhead{(mK)} &
  \colhead{($\rm 10^{19}$)} &
  \colhead{($\rm km\ s^{-1}$)} &
  \colhead{}\\
}

\startdata 

  1 & B2 2353+28+ & 00 00 14 & 28 19 16  &  110.181 & $-$33.065 &   21.0(2.0) &   25.7(3.3) & $-$116.2(1.3) &   2.3 &   3.2 & & &  H \\
  2 & 4C 49.48+  & 00 02 29 & 49 05 04  &  115.219 & $-$12.813 &  174.0(4.1) &   31.9(0.8) & $-$190.0(0.4) &   5.7 &   2.4 & & & H,A   \\
  3 & 4C 49.48+  & 00 02 29 & 49 05 04  &  115.219 & $-$12.813 &  &  &  &  &   2.4 & 0.27 &  $-$116 & H,A   \\
  4 & PKS 0003+15 & 00 03 25 & 15 53 08  &  107.318 & $-$45.325 &   17.0(2.1) &   29.8(4.2) & $-$326.9(1.8) &   2.8 &   2.4 & & &   H   \\
  5 & 4C 40.01$-$  & 00 04 02 & 40 34 58  &  113.861 & $-$21.220 &   13.3(1.4) &   33.0(1.0) & $-$402.0(0.4) &   1.9 &   2.4 & & &   H   \\
  6 & 4C 40.01$-$  & 00 04 02 & 40 34 58  &  113.861 & $-$21.220 &   29.0(0.0) &  100.0(0.0) &  308.0(0.0) &   0.0 &   2.4 & & &  UGC 00064   \\
  7 & PKS 0005$-$239 & 00 05 27 & $-$23 55 59 &   50.039 & $-$79.599 &  835.0(11.0) &   38.5(0.6) & $-$112.4(0.3) &  49.0 &  45.0 & & &   F   \\
  8 & PKS 0005$-$239+   & 00 12 19 & $-$23 55 59 &   53.723 & $-$81.042 &  570.0(10.0) &   47.1(1.0) & $-$123.3(0.4) &  48.0 &  43.0 & & &   F   \\
\enddata  
\tablecomments{Units of right ascension are hours, minutes, and
seconds, and units of declination are degrees, arcminutes, and
arcseconds. 
The quantities $\sigma_b$ and $\sigma_g$ are the rms noise 
of the data with   respect to 
 the polynomial baseline and the Gaussian, respectively.  
N$_{HI}$ and $<$V$>$ refer to emission in line wings ($\S3.1.1$).
The Notes indicate:  ``A''   Alternate reference position used; 
``F''  frequency-switched observations; ``H''  Hanning smoothed; ``U''  uncertain.}
\end{deluxetable} 

\begin{deluxetable}{rrrlrrrrl}
\tablecolumns{9}
\tablenum{4}
\tablewidth{0pt}
\tablecaption{High Velocity Gaussian Components}
\tablehead{
\colhead{} & \colhead{$\ell$} & \colhead{$b$} & 
\colhead{Name}  & \colhead{$\alpha$(B1950)} & \colhead{$\delta$(B1950)} &
\colhead{$V_{pk}$} & \colhead{$N_{HI}$} & \colhead{Complex} \\
\colhead{} & \colhead{(deg)} &  
\colhead{(deg)} & \colhead{} & \colhead{} & \colhead{}
& \colhead{(km s$^{-1}$)} & \colhead{(cm$^{-2}$)} &\colhead{}  \\
\colhead{(1)} &\colhead{(2)} &\colhead{(3)} &\colhead{(4)} &
\colhead{(5)} & \colhead{(6)} & \colhead{(7)} & \colhead{(8)} & \colhead{(9)}
}
\startdata
   1 &    5.451 & -32.289 & MS 20078-36+     &  20 14 46  &  -36 22 03   &     -114.4 &   3.36e+18 & GCN?  \\ 
   2 &   13.750 & -69.580 & PKS 2310-322+    &  23 17 20  &  -32 14 08   &     -109.3 &   1.25e+18 & GCN?  \\ 
   3 &   14.623 & -66.693 & PKS 2310-322-    &  23 03 35  &  -32 14 06   &     -159.9 &   6.69e+17 & GCN?  \\ 
   4 &   17.994 & -53.718 & PKS 2155-304+    &  22 02 36  &  -30 27 52   &     -115.1 &   1.30e+18 & GCN?  \\ 
   5 &   24.680 & -66.429 & PKS 2255-282+    &  23 02 15  &  -28 14 26   &     -155.0 &   1.01e+18 & GCN?  \\ 
   6 &   28.135 &  18.959 & PKS 1725+044+    &  17 32 48  &   04 29 30  &     -109.7 &   4.95e+18 & C  \\ 
   7 &   34.920 &  17.646 & OT 081           &  17 49 10  &   09 39 45  &     -139.6 &   7.21e+18 & C  \\ 
   8 &   34.920 &  17.646 & OT 081           &  17 49 10  &   09 39 45  &     -112.8 &   1.35e+19 & C  \\
\enddata

\tablecomments{Units of right ascension are hours, minutes, and
seconds, and units of declination are degrees, arcminutes, and
arcseconds. Values of $N_{HI}$ refer to the entire Gaussian 
even if part of it lies at $|V_{LSR}| < 100$ km s$^{-1}$.}

\end{deluxetable}

\vfill \eject

%

\begin{deluxetable}{ccccc}
\tablecolumns{5}
\tablenum{5}
\tablewidth{0pt}
\tablecaption{High-velocity Gaussian Components}
\tablehead{
\colhead{$N_{HI}$ range $(10^{18})$} & \colhead{Numb.\tablenotemark{a}}
 & \colhead{Percent\tablenotemark{b}}
& \colhead{$\Sigma N_{HI}\ (10^{20})$ \tablenotemark{c}} 
& \colhead{\% Total $N_{HI}$\tablenotemark{d}} \\
\colhead{(1)} &\colhead{(2)} &\colhead{(3)} &\colhead{(4)} &\colhead{(5)}
} 
\startdata
 0-1	&     42&   14	& 0.3 & 1.0 \\
 1-2	&     64&   21	& 0.9 & 2.9 \\
 2-3	&     34&   11	& 0.8 & 2.5 \\
 3-4	&     23&    7	& 0.8 & 2.5 \\
 4-5	&     17&    5	& 0.8 & 2.3 \\
 5-6	&     10&    3	& 0.5 & 1.7 \\
 6-8	&     16&    5	& 1.1 & 3.3 \\
 8-10	&     17&   5	& 1.5 & 4.5 \\
 10-15	&     22&   7	& 2.6 & 8.0 \\
 15-20	&     18&   6	& 3.0 & 9.3 \\
 20-30	&     16&   5	& 3.9 & 11.9 \\
 30-40	&     10&   3	& 3.6 & 10.9 \\
 40-50	&     5	&   2	& 2.1 & 6.5 \\
 50-75	&     10&   3	& 6.1 & 18.9 \\
 $>75$	&     5	&   2	& 4.5 & 14.2 \\
\enddata
\tablenotetext{a}{The number of detected high-velocity \hi\ lines in 
the particular $N_{HI}$ range.}
\tablenotetext{b}{The percentage of the total number of lines in the
particular $N_{HI}$ range.}
\tablenotetext{c}{The sum of the  \hi\ column density of all lines in the
given range of $N_{HI}$.}
\tablenotetext{d}{The percentage of the total observed N$_H$ contributed
by lines in this range of $N_{HI}$.}
\end{deluxetable}

\vfill \eject

\begin{deluxetable}{ccc}
\tablecolumns{3}
\tablenum{6}
\tablewidth{0pt}
\tablecaption{Average Velocity of  HI Lines}
\tablehead{
\colhead{Ref. Frame} & $<V_{em}>$ &
\colhead{$<V_{pk}>$ }  \\
\colhead{} & \colhead{(km s$^{-1}$)} & \colhead{(km s$^{-1}$)} \\
} 
\startdata
LSR & $-109\pm 120$ & $-118\pm 165$ \\
Gal Center & $-4 \pm 96 $ & $ -52 \pm 117$ \\
Local Group & $+15 \pm 106$ & $ -40\pm 121$ \\
\enddata

\tablecomments{The quantity 
$<V_{em}>$ is the average velocity weighted by the 
amount of emission at each velocity, i.e. by the value of $N_{HI}$.
The quantity  $<V_{pk}>$ is the average velocity of the line peaks, 
 regardless of their strength.}

\end{deluxetable}
\clearpage

\begin{deluxetable}{lrcccl}
\tablecolumns{6}
\tablenum{7}
\tablewidth{0pt}
\tablecaption{Angular Structure in Low-$N_{HI}$ Lines}
\tablehead{
\colhead{Source} & \colhead{$\ell$} &
\colhead{V$_{LSR}$} &
\colhead{$N_{HI}$($-$, S, +)}
& \colhead{$\Delta \theta$}
& \colhead{Complex}  \\
\colhead{} & \colhead{$(\deg)$} & \colhead{(km s$^{-1}$)} &
\colhead{ ($10^{18}$ cm$^{-2}$)} & \colhead{(deg)} 
} 
\startdata
MC1750+17 & 42 & $-120$ & 1.7, 9.0, 5.0 & 1.6 & C\\
4C 50.43  & 77 & $-130$ & 1.5, 6.8, 16.7 & 1.1 & C \\
PKS 2254+074 & 80 & $-330$ & 2.3, 8.3, 1.0 & 1.7 & MS \\
Mrk 205 & 124 & $-206$ & 2.8,  11.6,  0.7 & 0.9 & C \\
PG 1700+518 & 79 & $-170$ & 1.4, 1.3, 1.6 & 1.1 & C \\
B2 0051+29 & 122 & $-335$ & 1.2,  1.5,  2.0 & 1.5 & Wr  \\
MS03574+104 & 178 & $-240$ & 0.4,  3.4, 3.2 & 1.7 & ACVHV\\
PKS 1020-103 & 254 & $+160$ & 6.2,  1.4,  2.0 & 1.7 & WA \\
\enddata
\end{deluxetable}

\vfill \eject

\begin{deluxetable}{lc}
\tablecolumns{2}
\tablenum{8}
\tablewidth{0pt}
\tablecaption{Angular Structure in Triples}
\tablehead{
\colhead{Detections} & \colhead{Percent}
} 
\startdata
All Positions & 26 \\
Two Adjacent  & 16 \\
Edge Only & 49 \\
Center Only & 7 \\
All but Center & 2 \\
\enddata
\end{deluxetable}

\vfill \eject

\begin{deluxetable}{lrrrrl}
\tablecolumns{6}
\tablenum{9}
\tablewidth{0pt}
\tablecaption{Triples with \hi\  only at the Central Position}
\tablehead{
\colhead{Object} & \colhead{$\ell$} & \colhead{$b$} &
\colhead{V$_{LSR}$ } & \colhead{$N_{HI}$} & 
\colhead{Complex} \\
\colhead{} & \colhead{(deg)} & \colhead{(deg)} & \colhead{(km s$^{-1}$)} & 
\colhead{($10^{18}$)} \\ 
} 
\startdata 
PKS 2352-342      &   $359.941$ &  $-76.156$ &  172.3 &    2.0 & MS? \\
 UM 186            &    92.211 &  -60.618 & -312.5 &    1.5 & MS  \\
 PKS 1216-010      &   286.091 &   60.461 &  143.3 &    1.0   \\
 UM 301            &   128.892 &  -60.382 & -239.6 &    8.4   \\
 RXJ 21397+02      &    58.087 &  -35.006 & -223.1 &    0.5 & GCN  \\
 PKS 2254+074      &    80.425 &  -45.521 & -269.2 &    0.6 & MS  \\
 OT 081            &    34.920 &   17.646 & -112.8 &   13.5 & C  \\
 4C 09.72          &    86.450 &  -45.521 & -351.0 &    0.9 & MS \\
 PKS 0003+15       &   107.318 &  -45.325 & -326.9 &    1.0 & MS* \\
 Ton 34            &   202.957 &   56.507 &   98.9 &    1.8 &  \\
 B2 0051+29        &   123.532 &  -33.450 & -291.4 &    0.4 & Wr  \\
 B2 0051+29        &   123.532 &  -33.450 & -163.0 &    0.7 &  \\
\enddata
\end{deluxetable}

\vfill \eject

\end{document}